\documentclass[9pt,prr,twocolumn,superscriptaddress]{revtex4-2}

\usepackage[utf8]{inputenc}

\usepackage[T1]{fontenc}
\usepackage[french, english]{babel}
\usepackage{amsmath, amssymb, amsthm}
\usepackage{mathtools}
\usepackage{physics}
\usepackage{dsfont}
\usepackage{graphicx}
\usepackage{amsmath}
\usepackage{tabularx}   
\usepackage{placeins} 
\usepackage[dvipsnames]{xcolor}
\usepackage{float}

\usepackage{ragged2e}
\usepackage[compatibility=false,style=base]{caption}
\usepackage{subcaption} % une seule occurrence, après caption

\DeclareCaptionJustification{fulljustified}{\justifying}

\usepackage{media9}
\usepackage{animate}
\usepackage{tikz}
\usetikzlibrary{calc,arrows.meta, graphs, decorations.pathmorphing, positioning, decorations.pathreplacing, trees}{}
\usepackage{enumitem}
\usepackage[most]{tcolorbox}
\tcbuselibrary{theorems, breakable}

\usepackage[colorlinks=true, linkcolor=blue, citecolor=blue, urlcolor=blue]{hyperref}
\usepackage[nameinlink]{cleveref}

\usepackage{graphicx}
\usepackage{enumitem}
\usepackage{fmtcount}
\usepackage{diagbox}
\usepackage{overpic}
\usepackage{adjustbox}
\usepackage{relsize}
\usepackage[framemethod=TikZ]{mdframed} % Load mdframed with TikZ support
\usepackage{etoolbox} % Needed for \surroundwithmdframed
\RequirePackage{braket}
\usepackage{stmaryrd}
\usepackage{qcircuit}

\newcommand{\multimeter}[2]{%
    *+<1em,1.2em>{\hphantom{#2}} \POS [0,0]="i",[0,0].[#1,0]="e",  % Define box dimension
    !C + <0em,0.7em >*{#2}, % Insert the meter symbol in the box
    !C + <0em, -1.4 em>*{\metersymb}, % Properly position the label below
    "e"+UR;             % Drawing the box
        "e"+UL **\dir{-};
        "e"+DL **\dir{-};
        "e"+DR **\dir{-};
        "e"+UR **\dir{-},
    "i" \qw             % First qubit input qw
}

\newcommand{\hide}[1]{}

\begin{document}

% \tableofcontents

\title{Performance of Fusion-Based Quantum Computing with Imperfect Photon Indistinguishability}

\author{Katia Hakem}
\email{katia.hakem@quandela.com}
\affiliation{Quandela, 7 rue L\'eonard-de-Vinci, 91300 Massy, France}
\affiliation{LTCI, T\'el\'ecom Paris, Institut Polytechnique de Paris, 91120 Palaiseau, France}
\author{Gr\'egoire de Gliniasty}
\affiliation{Quandela, 7 rue L\'eonard-de-Vinci, 91300 Massy, France}
\author{Nathan Shettell}
\affiliation{MajuLab, CNRS-UCA-SU-NUS-NTU International Joint Research Laboratory and
Centre for Quantum Technologies, National University of Singapore, 117543 Singapore, Singapore}
\author{Paul Hilaire}
\affiliation{LTCI, T\'el\'ecom Paris, Institut Polytechnique de Paris, 91120 Palaiseau, France}
\author{Stephen C. Wein}
\affiliation{Quandela, 7 rue L\'eonard-de-Vinci, 91300 Massy, France}

\date{\today}

\begin{abstract}

Fusion-based quantum computation (FBQC) is a promising route to fault-tolerant photonic quantum computing, connecting small entangled resource states through linear-optical fusion measurements. While logical encoding can substantially improve resilience to photon loss, whether these benefits extend to other noise remain an open question.
In this work, we show that this is not generally the case when accounting for partial photon distinguishability during resource state generation and fusion measurement. We develop hardware-motivated noise models to capture these effects and investigate their impact on fault tolerance in the 6-ring fusion network. We compare unencoded and Shor\((2,2)\)-encoded resource states prepared using all-photonic and emitter-assisted repeat-until-success (RUS) schemes. For the all-photonic construction, encoding reduces the distinguishability threshold from \(0.136\%\) to \(0.080\%\). In contrast, the RUS construction tolerates more distinguishability, with comparable unencoded and encoded thresholds (0.256\% and 0.271\%, respectively). These results expose a fundamental limitation of evaluating logical encodings under idealized resource-state assumptions and demonstrate that resource-state generation can critically determine the fault tolerance of photonic architectures.

\end{abstract}

\maketitle

\section{Introduction}

Building a scalable, fault-tolerant quantum computer requires understanding how hardware imperfections affect error correction performance. Among the platforms pursued toward this goal, photonics is attractive because photons propagate rapidly and interact weakly with their environment. However, their weak
interactions make entangling operations difficult to implement~\cite{LOQC_review}. In photonic measurement-based approaches, computation proceeds through single-qubit measurements on a large entangled resource state~\cite{MBQC}. Preparing such states with high fidelity is challenging because linear-optical entangling operations are probabilistic~\cite{Nielsen2004,Browne2005}. 

Fusion-based quantum computation (FBQC) addresses this challenge using small, constant-size entangled resource states and destructive entangling measurements, called fusions~\cite{Browne2005,FBQC}. Products of fusion outcomes define parity checks for error correction and decoding~\cite{FBQC}. Evaluating the fault tolerance of a fusion network therefore requires determining how physical imperfections in resource-state preparation and fusion affect those outcomes.

The resource states required for FBQC can be assembled from photons produced by spontaneous parametric down-conversion~\cite{SPDC} or four-wave mixing~\cite{MeyerScott2020}, using probabilistic linear-optical operations combined with heralding and multiplexing. Alternatively, quantum emitters such as atoms~\cite{Thomas2022} and quantum dots~\cite{Senellart2017,Schwartz2016} can generate photonic states through sequential spin–photon entanglement. A single emitter can generate linear cluster states~\cite{Lindner2009}, and more generally caterpillar states~\cite{huet2025deterministic,Hilaire2023neardeterministic} (which include GHZ states) while multiple interacting emitters can generate more general photonic graph states~\cite{Li2022,GimenoSegovia2019}.

Photon loss is a leading hardware limitation in photonic architectures and has consequently been a primary focus of FBQC analyses~\cite{FBQC,bartolucci2025comparison,bombin2023fault, bombin2023increasing,Macromux,litinski2025blocklet,bell2023optimizing,PaesaniBrown2023,lobl2026subthreshold,Song2024}. For dual-rail photonic qubits with photon-number-resolving (PNR) detectors, photon loss changes the detected photon number, whereas intrinsic fusion failure produces a known detection pattern. Both events are therefore heralded and can be represented as erasures of the fusion outcomes. Another imperfection is partial photon distinguishability, which arises when photons do not perfectly overlap in their spectral, temporal, or spatial modes. This imperfection reduces the visibility of two-photon interference thereby retaining residual which-way information~\cite{HOM,vanDenHoven2026}. The resulting degradation of multiphoton interference can produce an accepted detection pattern associated with an incorrect fusion parity. These errors are unheralded and can affect fusion outcomes directly or generate correlated errors during resource-state preparation, potentially leading to logical failure.

Threshold studies have explored photon loss and fusion failure across a range of FBQC architectures~\cite{FBQC,PaesaniBrown2023, Song2024,bombin2023increasing,Macromux}. The original FBQC analysis describes unheralded faults using a hardware-agnostic model of independent fusion-outcome
flips~\cite{FBQC}. Relating these error probabilities to photon distinguishability requires determining how imperfect interference affects both fusion measurements and the preparation of their input resource states. In architectures that assemble resource states through successive interference and postselection stages, preparation errors can propagate into several fusion outcomes and generate correlations.

In this work, we derive optical circuit-dependent
distinguishability error channels for fusion measurements and resource-state preparation, then incorporate them into fault-tolerance simulations of the unencoded six-ring fusion network and its Shor$(2,2)$-encoded counterpart. For the encoded fusion network, we study the all-photonic and emitter-assisted repeat-until-success (RUS) preparation schemes proposed by Wein \textit{et al.}~\cite{Wein2025}. Although these schemes produce the same ideal encoded resource state, their distinct preparation circuits lead to different effective error channels. This motivates the choice of preparation scheme to enhance the distinguishability threshold.

The paper is organized as follows. Section~\ref{sec:Background} reviews the theoretical background on photonic fusion-based quantum computation.
Section~\ref{sec:noise-model} introduces the distinguishability
model and the resulting fusion error channels.
Sections~\ref{sec:unencoded-six-ring}
and~\ref{sec:shor_22_encoded_six_ring} describe the preparation
of unencoded and Shor$(2,2)$-encoded six-ring resource states,
respectively, considering both all-photonic and emitter-assisted
RUS schemes. Section~\ref{sec : Hardware-motivated FBQC simulation}
presents the fault-tolerance simulations and threshold estimates.
Section~\ref{sec:Discussion} discusses the results and their
limitations, and Section~\ref{sec:Conclusion} concludes
with an outlook.

\section{Background}
\label{sec:Background}
This section reviews the key concepts of photonic fusion-based quantum computation needed to understand and interpret the results of our work.
\begin{figure}[h!]
\centering
\begin{subfigure}[b]{0.45\linewidth}
    \centering
    \caption*{\centering\textbf{ {$\left| \overline{0} \right\rangle \equiv \left|1, 0\right\rangle_{1,2} = a_1^\dagger \ket{0,0}_{1,2}$}}}
    \begin{tikzpicture}[scale=1, photon/.style={draw, decorate, decoration={snake, amplitude=3pt, segment length=7pt}, thick}]
        
    % Voies 1 et 2 pour |0⟩
    \draw[thick] (-1,2) -- (1,2); % voie 1
    \draw[thick] (-1,1.5) -- (1,1.5); % voie 2
    
    % Photon sur voie 1
    \draw[photon, blue] (-0.2,2) -- (0.2,2);
    \node[right] at (1.1,1.75){};
    \end{tikzpicture}
\end{subfigure}
\hfill
\begin{subfigure}[b]{0.45\textwidth}
    \centering
    \caption*{\centering\textbf{ {$\left| \overline{1} \right\rangle \equiv \left|0, 1\right\rangle_{1,2} = a_2^\dagger \ket{0,0}_{1,2}$}}}
        \begin{tikzpicture}[scale=1, photon/.style={draw, decorate, decoration={snake, amplitude=3pt, segment length=7pt}, thick}]

    % Voies 1 et 2 pour |1⟩
\draw[thick] (3,2) -- (5,2); % voie 1
\draw[thick] (3,1.5) -- (5,1.5); % voie 2

% Photon sur voie 2
\draw[photon, blue] (3.8,1.5) -- (4.2,1.5);
\node[right] at (5.1,1.75){};
\end{tikzpicture}
\end{subfigure}
\caption{Dual-rail encoding of a photonic qubit in two optical modes. The blue single photon indicates the occupied mode, while the other mode is empty. A valid dual-rail state contains exactly one photon across the two modes.}
\label{fig:dual-rail}
\end{figure}
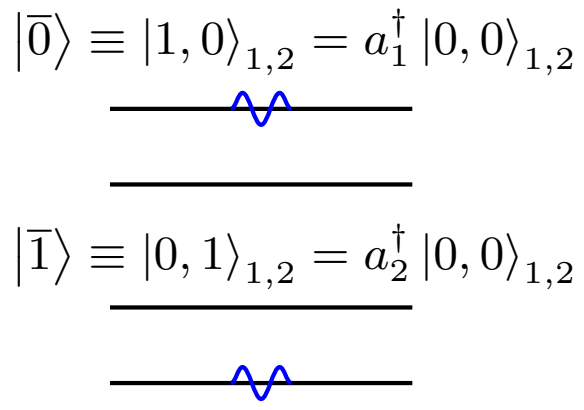
\subsection{Dual-rail encoding }
\label{sec:dual-rail}
We consider dual-rail photonic qubits, where a single photon occupies one of two optical modes. The states $\ket{{0}}$ and $\ket{{1}}$ correspond to occupation of the first and second rail, respectively, as illustrated in Fig.~\ref{fig:dual-rail}. Here, $\hat{a}_i^\dagger$ denotes the creation operator associated with optical mode $i$, with $[\hat{a}_i,\hat{a}_j^\dagger]=\delta_{ij}$. A photon-loss event maps the qubit to the vacuum state \(\ket{0,0}_{1,2}\), which lies outside the logical subspace and represents an erasure.

For perfectly indistinguishable photons, we assume identical temporal, spectral, and polarization states; the spatial modes encode the qubits. This assumption is relaxed in Sec.~\ref{sec:noise-model}, where partial photon distinguishability is introduced.

\subsection{Graph states}
\label{sec:resource-states}

In the FBQC schemes considered here, the computation is built from small, constant-size entangled stabilizer states, referred to as resource states~\cite{FBQC}. Such states can be represented, up to local Clifford transformations, as graph states~\cite{Hein2004}. A graph state is associated with a graph $G=(V,E)$  with vertex set $V$ and edge set $E$. Each vertex is associated with a qubit prepared in the $\ket{+}=(\ket{0}+\ket{1})/\sqrt{2}$ state, and each edge corresponds to a $CZ$ gate on the corresponding qubits:
\begin{equation}
\ket{G}=\prod_{(i,j)\in E} CZ_{i,j}\ket{+}^{\otimes |V|}.    
\end{equation}

Equivalently, $\ket{G}$ is the unique simultaneous $+1$ eigenstate of
the $ \abs{V}$ independent stabilizer generators
\begin{equation}
    K_i
    =
    X_i\prod_{j\in\mathcal{N}(i)}Z_j,
    \qquad
    K_i\ket{G}=\ket{G},
    \qquad i\in V,
    \label{eq:graph-stabilizers}
\end{equation}
where $X_i$ and $Z_i$ are single-qubit Pauli operators and $\mathcal{N}(i)$ denotes the neighborhood of vertex $i$.

\begin{figure}[t!]
    \centering
    \begin{tikzpicture}[
        vertex/.style={
            circle,
            draw=teal!75!black,
            fill=teal!25,
            minimum size=5mm,
            inner sep=0pt,
            font=\scriptsize,
            line width=0.8pt
        },
        central/.style={vertex, draw=red!30!black, fill=red!45},
        neighbor/.style={vertex, draw=blue!70!black, fill=blue!25},
        edge/.style={draw=teal!75!black, line width=0.9pt}
    ]

    \def\radius{1.15}

    % Symmetric bounding box: the ring sits at the horizontal centre of the figure
    \path[use as bounding box] (-2.3,-3.15) rectangle (2.3,1.7);

    % ---------------- Six-ring ----------------
    \node[central]  (v1) at (180:\radius) {1};
    \node[neighbor] (v2) at (120:\radius) {2};
    \node[vertex]   (v3) at (60:\radius)  {3};
    \node[vertex]   (v4) at (0:\radius)   {4};
    \node[vertex]   (v5) at (-60:\radius) {5};
    \node[neighbor] (v6) at (-120:\radius){6};

    \draw[edge] (v1)--(v2)--(v3)--(v4)--(v5)--(v6)--(v1);

    % Pauli operators entering K_1
    \node[font=\scriptsize, text=red] at (-1.75,0) {$X_1$};
    \node[font=\scriptsize, text=blue!70!black] at (-0.95,1.35) {$Z_2$};
    \node[font=\scriptsize, text=blue!70!black] at (-0.95,-1.35) {$Z_6$};

    % Example stabilizer
    \node[font=\small] at (0,-1.8)
        {$K_1=\textcolor{blue}{Z_6}\textcolor{red}{X_1}\textcolor{blue}{Z_2}$};

    \draw[edge] (v1)--(v2)--(v3);
\draw[edge] (v3) -- node[midway, right=3pt, font=\scriptsize] {$CZ$} (v4);
\draw[edge] (v4)--(v5)--(v6)--(v1);
\node[font=\scriptsize, right=3pt] at (v4.east) {$\ket{+}$};
    \end{tikzpicture}
    \vspace{-3em}
    \caption{Six-ring graph state used as a resource state in the fusion network. Each vertex represents a photonic qubit initialized in the $\ket{+}$ state and each edge represents a $CZ$ operation. The highlighted vertices illustrate the stabilizer generator associated with vertex \(1\), \(K_1=Z_6X_1Z_2\).}
    \label{fig:six-ring-resource}
\end{figure}
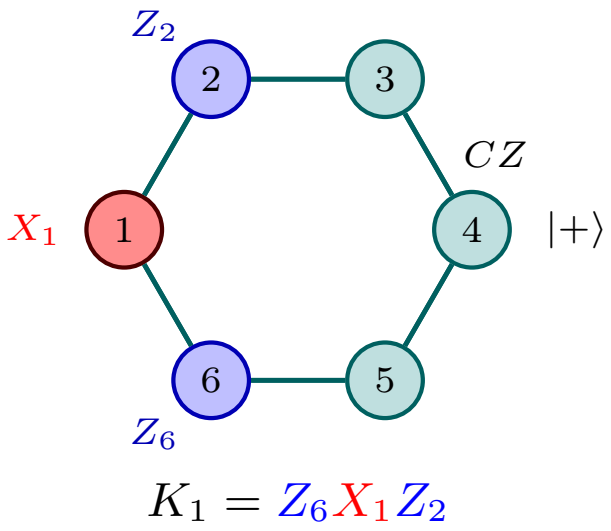

One of the resource states considered in this work is the six-ring graph state shown in Fig.~\ref{fig:six-ring-resource}. Since every vertex has two neighbors, its stabilizer generators take the form
\begin{equation}
    K_i=Z_{i-1}X_iZ_{i+1}.
\end{equation}

\subsection{Resource state generation}

We now discuss how photonic resource states can be prepared, considering two families of photon sources.

\paragraph{Heralded single-photon sources.}
These sources rely on nonlinear optical processes such as spontaneous parametric down-conversion (SPDC)~\cite{SPDC}, which generate photon pairs probabilistically: the detection of one photon heralds the presence of its partner. Entangled states are then built by fusing the heralded photons. Because every fusion succeeds only probabilistically, the success probability of generating a multi-photon entangled state decreases rapidly with its size. Compensating for this requires substantial resource overhead, typically through multiplexing.

\paragraph{Quantum emitters.}
Here, a quantum emitter with a spin qubit sequentially emits photons that are entangled with the spin, which enables the deterministic generation of linear photonic graph states~\cite{Lindner2009,coste2023high}. This approach avoids the probabilistic generation of the initial photons, but it requires the emitter to remain coherent throughout the emission sequence to produce high-fidelity entangled states.

In the following section, these two approaches are illustrated  through the generation of the six-ring resource state, using respectively a purely photonic scheme and a scheme based on six sequentially controlled quantum emitters.
\subsection{Fusion gates}
\label{sec:Fusion}

\begin{figure}[ht]
    \centering
    \begin{tikzpicture}[
        graph/.style={
            circle,
            draw=red!60!black,
            fill=purple!20,
            minimum size=5.8mm,
            thick,
            inner sep=0pt
        },
        textnode/.style={
            rectangle,
            draw=none,
            fill=none,
            inner sep=1pt,
            font=\footnotesize
        },
        thick
    ]

        % First three-qubit GHZ graph state
        \node[graph] (a1) at (0,0.65) {1};
        \node[graph] (a2) at (0,-0.65) {2};
        \node[graph] (a3) at (1,0) {3};
        \draw (a1) -- (a3);
        \draw (a2) -- (a3);

        % Second three-qubit GHZ graph state
        \node[graph] (b4) at (2,0) {4};
        \node[graph] (b5) at (3,0.65) {5};
        \node[graph] (b6) at (3,-0.65) {6};
        \draw (b4) -- (b5);
        \draw (b4) -- (b6);

        \draw[
            red!60!black,
            dashed,
            rounded corners=3pt,
            line width=0.9pt
        ]
        ($(a3)+(-0.35,0.38)$)
        rectangle
        ($(b4)+(0.35,-0.38)$);

        % Type-II branch
        \draw[->] (3.45,0.15) -- (5.15,1.20);
        \node[textnode, align=center] at (4.45,1.55)
        {Type-II: \\[-0.5mm] \(\{X_3X_4,Z_3Z_4\}\)};

        % Four-qubit GHZ graph state
        \node[graph] (c1) at (6.0,1.80) {1};
        \node[graph] (c2) at (6.0,0.60) {2};
        \node[graph] (c5) at (7.0,1.20) {5};
        \node[graph] (c6) at (8.0,1.20) {6};
        \node[textnode] at (7.0,1.70) {$H_5$};
        \draw (c1) -- (c5);
        \draw (c2) -- (c5);
        \draw (c5) -- (c6);

        % Type-I branch
        \draw[->] (3.45,-0.15) -- (5.15,-1.20);
        \node[textnode, align=center] at (4.45,-1.55)
        {Type-I: \(\{Z_3Z_4\}\)};

        % Five-qubit star graph resulting from Type-I fusion
        \node[graph] (d1) at (6.0,-0.60) {1};
        \node[graph] (d2) at (6.0,-1.80) {2};
        \node[graph] (df) at (7.0,-1.20) {\(f\)};
        \node[graph] (d5) at (8.0,-0.60) {5};
        \node[graph] (d6) at (8.0,-1.80) {6};

        \draw (d1) -- (df);
        \draw (d2) -- (df);
        \draw (df) -- (d5);
        \draw (df) -- (d6);

    \end{tikzpicture}
    \caption{Action of successful fusion operations on two three-qubit star graph states, each locally Clifford equivalent to a GHZ state. The dashed region identifies the input qubits \(3\) and \(4\). Type-II fusion jointly measures \(X_3X_4\) and \(Z_3Z_4\) and consumes both input qubits. The resulting state is locally Clifford equivalent to the four-qubit star graph shown, with \(H_5\) denoting the corresponding local Hadamard. Type-I fusion measures the observable \(Z_3Z_4\) and maps the two input qubits onto the output qubit \(f\), which inherits their combined neighborhood. Pauli corrections are omitted.}
    \label{fig:fusion}
\end{figure}
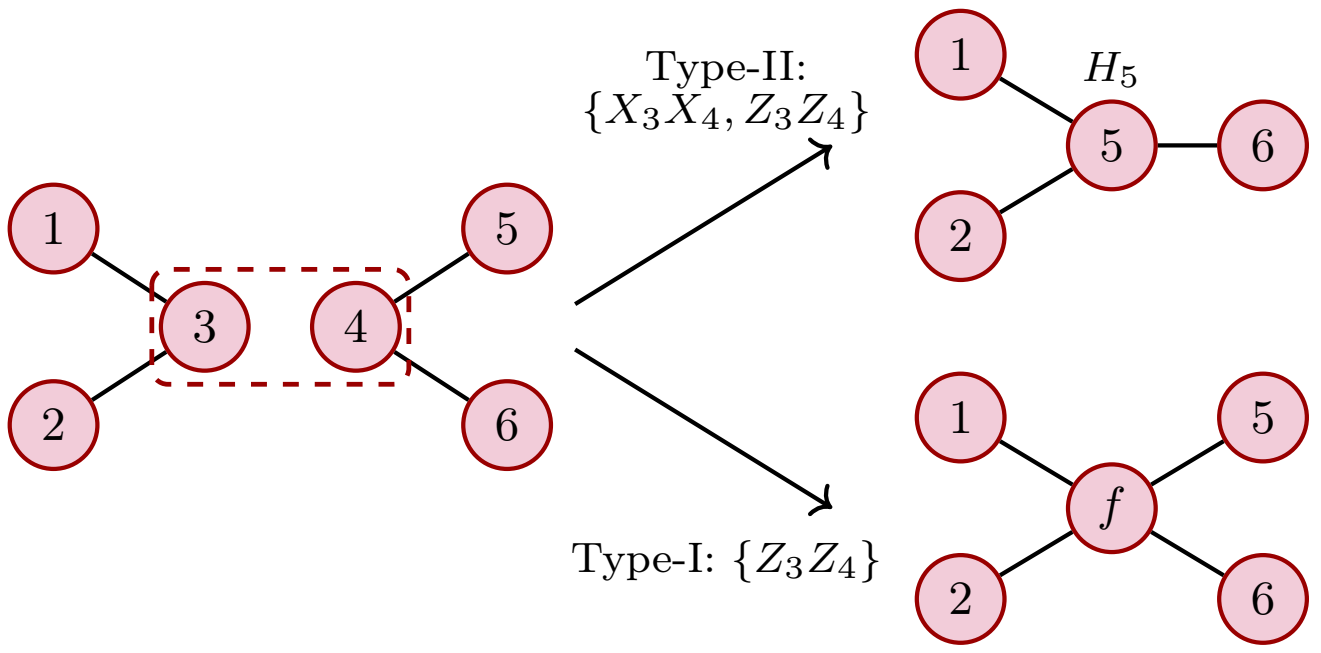

In the stabilizer formalism, a fusion operation can be described in terms of parity measurements on two input qubits~\cite{Browne2005}.
A Type-II fusion is a destructive joint measurement of the two observables \(XX\) and \(ZZ\). The pair of outcomes $(XX,ZZ)$ uniquely identifies the four Bell states, so the operation performs a Bell-state measurement on the two input qubits. Both input qubits are consumed, while the qubits in their neighborhoods are projected onto an entangled state. By contrast, a successful Type-I fusion performs a single parity projection, taken here to be \(ZZ\), and maps the two input qubits onto a single surviving output qubit. Examples of successful Type-I and Type-II fusions are illustrated in Fig.~\ref{fig:fusion}.

In the absence of noise, standard linear-optical Type-I and Type-II fusions succeed with probability \(1/2\). A failed Type-II fusion reduces to separate \(Z\)-basis measurements of the two input qubits, up to local basis changes. Consequently, the \(XX\) outcome is erased, whereas the \(ZZ\) parity remains available as the product of the two individual measurement outcomes. The parity information associated with each possible Type-II fusion outcome is summarized in
Table~\ref{tab:bell_fusion_outcomes}.

\begin{table}[H]
\centering
\begin{tabular}{|c||c|}
\hline
\textbf{Outcome} & \textbf{Available parity information} \\
\hline
Success & \(XX,\;ZZ\) \\
Intrinsic failure & \(ZZ\) \\
erasure (loss)& None \\
\hline
\end{tabular}
\caption{Parity information associated with the possible outcomes of a Type-II fusion, using the \(Z\)-basis failure convention.}
\label{tab:bell_fusion_outcomes}
\end{table}

The intrinsic fusion-failure probability can be reduced using additional photons. In the scheme introduced by Grice~\cite{Grice2011},
an auxiliary state containing \(2^n-2\) photons reduces the Type-II fusion-failure probability to
\begin{equation}
    p_{\mathrm{fail}}=2^{-n}.
\end{equation}
For \(n=2\), an auxiliary Bell-pair qubit therefore yields \(p_{\mathrm{fail}}=1/4\). In this work, we instead employ the Ewert \& van Loock construction~\cite{3/4BSM}, which achieves the same \(75\%\) success probability using four ancillary single photons. Type-I fusion can similarly be boosted from \(1/2\) to \(3/4\) using a permuted dual-rail Bell-state ancilla~\cite{Bartolucci2021Creation}.

\subsection{Repeat-until-success gates}

Boosted gates address the challenge of probabilistic entangling gates in linear optics by adding additional auxiliary photons. These extra resources are consumed during a single attempt, and a failure generally terminates the operation. In contrast, repeat-until-success (RUS) gates~\cite{Lim2005} use ancillary photons to herald the outcome of an entangling operation performed on emitter qubits. When a failure outcome is obtained, the resulting state of the emitter qubits is known and can be corrected using single-qubit operations or a Pauli-frame update. The same operation can then be attempted again~\cite{Lim2005,Bose1999}. If the success probability of one RUS attempt is \(p_{\mathrm{s}}\), the probability of obtaining at least one successful outcome after \(n\) attempts is
\begin{equation}
P_{\mathrm{s}}(n)=1-(1-p_{\mathrm{s}})^n .
\end{equation}
The fusion success probability \(P_{\mathrm{s}}(n)\) approaches
unity as the boosting level \(n\) tends to infinity. In practice, the increase of \(P_{\mathrm{s}}(n)\) is limited by is limited by photon loss, photon distinguishability and the decoherence of the stationary qubits. The two approaches therefore have different types of overhead: boosted gates introduce additional resource overhead, whereas RUS gates introduce temporal overhead through repeated attempts.

In this work, photons emitted by two emitters interfere in a linear-optical interferometer and their detection heralds a RUS $CZ$ operation between the emitter spins as introduced in Ref.~\cite{SPOQC}. These RUS $CZ$ gates can thus generate the edges of the six-ring resource state without requiring a direct interaction between the spin qubits.

\section{Hardware-motivated noise model}
\label{sec:noise-model}

The goal of fault-tolerant fusion networks is to protect quantum information from physical errors such as  intrinsic fusion failure, photon loss, and partial photon distinguishability. These errors have different effects: fusion failure and photon loss are heralded, whereas distinguishability can introduce unheralded errors in the reported parity outcomes. Reference~\cite{FBQC} considers a hardware-agnostic model in which each \(XX\) or \(ZZ\) fusion outcome is independently flipped with probability \(p_{\mathrm{error}}\) or erased with probability
\(p_{\mathrm{erasure}}\).  By relating the erasure probability \(p_{\mathrm{erasure}}\) to photon loss and fusion failure, they obtain fault-tolerance loss thresholds. However, the error parameter $p_{\mathrm{error}}$ remains abstract and is not derived from a specific physical imperfection.  Here, we derive circuit-dependent error models that connect partial photon distinguishability to fusion-outcome errors and imperfections introduced during resource-state generation. These models are incorporated as error channels in our simulation of the fusion-network circuit.

\subsection{Photon loss and partial distinguishability}

Each photon is independently lost with probability \(p_{\mathrm{loss}}\). Assuming ideal photon-number-resolving (PNR) detectors, loss events are heralded. At the fusion level, such an event is treated as a full erasure of both observables $XX$ and $ZZ$. Intrinsic fusion failure is also heralded but one parity outcome can be recovered ($ZZ$ in our convention) as discussed in Sec.~\ref{sec:Fusion}.

For two photons with normalized internal states
\(\ket{\phi_i}\) and \(\ket{\phi_j}\), we define their
distinguishability as
\begin{equation}
D=1-\left|\braket{\phi_i|\phi_j}\right|^2.
\end{equation}
Thus, \(D=0\) corresponds to perfectly indistinguishable
photons, whereas \(D=1\) corresponds to orthogonal internal
states. For pure single-photon inputs interfering on an
ideal balanced beam splitter, \(D=1-V_{\mathrm{HOM}}\),
where \(V_{\mathrm{HOM}}\) is the Hong-Ou-Mandel
visibility~\cite{HOM}.

We describe multiphoton distinguishability using the
orthogonal bad-bit (OBB) model~\cite{sparrow2018quantum}.
Let \(\hat a_{r,\mu}^\dagger\) create a photon in spatial
mode \(r\) and internal mode \(\mu\), with
\begin{equation}
[\hat a_{r,\mu},\hat a_{s,\nu}^\dagger]
=
\delta_{rs}\delta_{\mu\nu}.
\end{equation}
The internal mode \(\mu=0\) is common to all photons,
while each photon \(j\) has its own orthogonal bad mode
\(\mu=j\). Its creation operator is
\begin{equation}
\hat c_{r,j}^\dagger
=
\sqrt{1-\epsilon}\,\hat a_{r,0}^\dagger
+
\sqrt{\epsilon}\,\hat a_{r,j}^\dagger,
\label{eq:obb-creation-operator}
\end{equation}
where \(\epsilon\) is the single-photon bad-mode
probability. Since the bad modes are mutually orthogonal,
the internal-state overlap between any two distinct
photons is \(1-\epsilon\). Their common HOM visibility
therefore satisfies
\begin{equation}
V_{\mathrm{HOM}}=(1-\epsilon)^2.
\end{equation}
Consequently,
\begin{equation}
\epsilon
=
1-\sqrt{V_{\mathrm{HOM}}}
=
1-\sqrt{1-D}
=
\frac{D}{2}+\mathcal O(D^2).
\label{eq:obb-visibility-relation}
\end{equation}
Equivalently, Eq.~\eqref{eq:obb-creation-operator} can be
written in the form used in Ref.~\cite{sparrow2018quantum},
\begin{equation}
\hat c_{r,j}^\dagger
=
V_{\mathrm{HOM}}^{1/4}\hat a_{r,0}^\dagger
+
\sqrt{1-\sqrt{V_{\mathrm{HOM}}}}\,
\hat a_{r,j}^\dagger.
\end{equation}

The interferometer preserves the internal modes, and
the detectors do not resolve them. For \(N\) photons
with independent errors, the all-good
configuration has probability \((1-\epsilon)^N\), while
each configuration containing exactly one bad photon has
probability \(\epsilon(1-\epsilon)^{N-1}\).
To first order in \(D\), this gives the input state $\rho_{\mathrm{in}}$: 
\begin{equation}
\rho_{\mathrm{in}}
=
\left(1-\frac{ND}{2}\right)\rho_0
+
\frac{D}{2}\sum_{j=1}^{N}\rho_j
+
\mathcal O(D^2).
\label{eq:obb-input-state}
\end{equation}
Here, \(\rho_0\) is the ideal input state with all photons
in internal mode \(0\), and \(\rho_j\) is the same input
with photon \(j\) in its orthogonal internal mode \(j\). Two or more bad photons contribute at order \(D^2\) or higher.
\subsection{Unboosted fusion operations}
\label{sec:unboosted}

\begin{figure}[ht]
    \centering
    \input{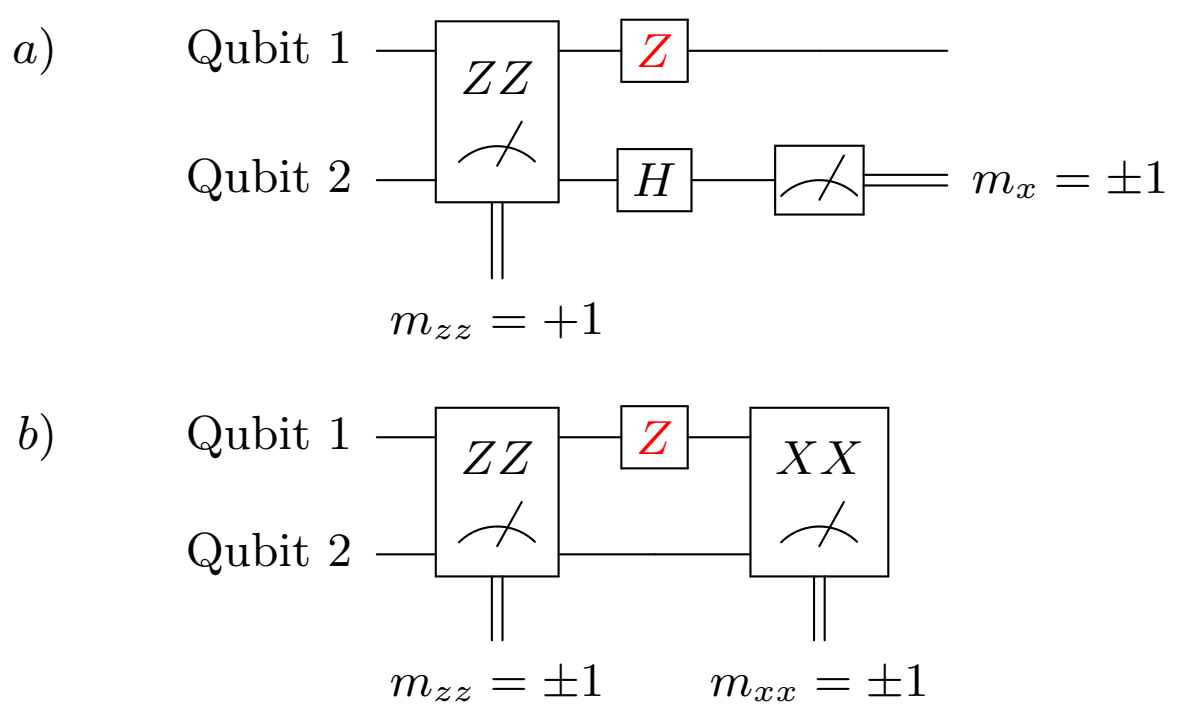}

    \caption[Effective noise models for unboosted fusions]{%
    \justifying
    Effective circuits and distinguishability noise models for
    (a) a successful unboosted Type-I fusion and
    (b) a successful unboosted Type-II fusion.
    Partial photon distinguishability is represented by a
    stochastic \(Z\) error (red) after the \(ZZ\) measurement,
    occurring with probability \(D/2\).
    For Type-I fusion, it produces a phase flip on the surviving
    photonic qubit. For Type-II fusion, it flips the reported
    outcome \(m_{xx}\), while leaving \(m_{zz}\) unchanged.
    }
    \label{fig:fusion_circuits}
\end{figure}

The resulting noise models for unboosted Type-I and Type-II fusions are shown in Fig.~\ref{fig:fusion_circuits}, respectively. Conditioned on a successful Type-I fusion, partial photon distinguishability produces a phase-flip error $Z$ on the surviving qubit $f$ with probability $D/2$. For a Type-II fusion, it flips the reported outcome $m_{xx}$ of the $XX$ parity measurement with probability $D/2$, while the $ZZ$ outcome $m_{zz}$ is unaffected, as derived in Ref.~\cite{sparrow2018quantum}.

\subsection{Boosted Type-II fusion}

To reduce the intrinsic fusion-failure probability, we consider two boosted Type-II fusion schemes based on the construction of Ewert and van Loock (EL)~\cite{3/4BSM}. We choose this approach because its first level uses initially unentangled single photons, avoiding errors induced by the preparation of an auxiliary Bell pair required by the Grice scheme used in Ref.~\cite{FBQC}. 
The first-level scheme uses four auxiliary single photons
and increases the ideal fusion success probability from \(1/2\)
to \(3/4\). The second level adds two four-photon entangled
ancillary states, giving twelve auxiliary photons and an ideal
success probability of \(7/8\). For this second level, we apply
distinguishability noise directly to the entangled ancillary
states without assuming a specific preparation circuit.
Our schemes use more photons than the corresponding Grice
constructions and are therefore more sensitive to photon
loss~\cite{3/4BSM,Grice2011}.

In general, partial distinguishability can cause both \emph{unheralded parity flips} and \emph{heralded parity erasures}. A detection pattern that ideally identifies one Bell state can arise from another input under noise, producing an incorrect reported parity. Distinguishability also enables patterns that are that are never observed in the ideal case. Some still determine \(ZZ\), while others erase both parity outcomes. These full erasures differ from intrinsic fusion failure, which erases only \(XX\) while retaining \(ZZ\) in our convention.

Within the first-order OBB model, we include distinguishability
of all input and auxiliary photons, using
\(D=1-V_{\mathrm{HOM}}\).
We assume an ideal interferometer and PNR detectors that do not resolve internal modes.
Table~\ref{tab:boosted-fusion-outcomes} summarizes the resulting outcome probabilities in the absence of photon loss.
Applying Hadamard operations to both input qubits interchanges
the roles of \(XX\) and \(ZZ\).

\begin{table}[ht]
\centering
\caption{Outcome probabilities for boosted Type-II fusion,
averaged over equiprobable Bell inputs, in the absence of
photon loss.}
\label{tab:boosted-fusion-outcomes}
\renewcommand{\arraystretch}{1.25}
\begin{tabular}{lcc}
\hline
Outcome & \(3/4\) scheme & \(7/8\) scheme \\
\hline
Both parities correct
& \(\frac34-\frac{5D}{4}\)
& \(\frac78-\frac{77D}{32}\) \\
Only \(XX\) flipped
& \(\frac{3D}{8}\)
& \(\frac{23D}{32}\) \\
Only \(ZZ\) flipped
& \(\frac{3D}{16}\)
& \(\frac{3D}{16}\) \\
Both parities flipped
& \(\frac{3D}{16}\)
& \(\frac{3D}{16}\) \\
Only \(XX\) erased
& \(\frac14-\frac{3D}{16}\)
& \(\frac18+\frac{5D}{16}\) \\
Both parities erased
& \(\frac{11D}{16}\)
& \(D\) \\
\hline
\end{tabular}
\end{table}

A fusion is reported as successful when both parity outcomes are reported, even if either is incorrect. Conditioned on success, the probability of at least one parity flip is \(D\) for the \(3/4\) scheme and \(5D/4\) for the \(7/8\) scheme, to first order. Second-level boosting therefore halves the intrinsic failure probability while increasing both the conditional parity-error probability and the probability of a full erasure. The errors also become more strongly biased toward \(XX\) flips, corresponding to effective \(Z\) errors on an input qubit. Detailed derivations and the success-conditioned Pauli channels are given in Appendices~\ref{app:boosted-fusion-noise} and~\ref{app:boosted-fusion-noise-7-8}, respectively.

\subsection{Repeat-until-success CZ gate}
\label{sec:rus-noise}
For the emitter-based architecture, we adopt the effective distinguishability channel for a successful RUS CZ operation derived in Ref.~\cite{SPOQC},
\begin{equation}
\mathcal{C}_{D}(\rho)
=
\left(1-\frac{D}{2}\right)
\mathrm{CZ}\rho\mathrm{CZ}
+
\frac{D}{2}
Z_1Z_2\,\mathrm{CZ}\rho\mathrm{CZ}\,Z_2Z_1 .
\label{eq:rus-distinguishability-channel}
\end{equation}

The RUS CZ interferometer is similar to that used for Type-II fusion. As in Fig.~\ref{fig:fusion_circuits}, partial photon distinguishability can flip a reported measurement outcome. Here, this outcome determines whether the operation applied to the spin qubits is $\mathrm{CZ}$ or $Z_1Z_2\mathrm{CZ}$, and hence which Pauli correction is required; see Table~1 of Ref.~\cite{SPOQC}. An incorrect outcome therefore leaves a residual $Z_1Z_2$ error on the spins. Since repeat outcomes are unaffected by distinguishability, these errors do not accumulate over repeated attempts, and Eq.~\eqref{eq:rus-distinguishability-channel} describes the entire RUS gate conditioned on eventual success.

\section{Unencoded six-ring fusion network}
\label{sec:unencoded-six-ring}
The six-ring fusion network introduced in Ref.~\cite{FBQC} consists of six-qubit ring graph states, shown in
Fig.~\ref{fig:six-ring-resource}, arranged in a three-dimensional network with two resource states per cubic unit cell. Type II fusions connect qubits from neighboring resource states by measuring their \(XX\) and \(ZZ\) parities. Reference~\cite{FBQC} characterized this network using a hardware-agnostic noise model in which individual fusion outcomes are independently erased or flipped. The marginal thresholds are \(p_{\mathrm{erasure}}^{\mathrm{th}}=11.98\%\) and \(p_{\mathrm{error}}^{\mathrm{th}}=1.07\%\). In the absence of photon loss and parity errors, uniformly randomizing the fusion failure basis gives \(p_{\mathrm{erasure}}=p_{\mathrm{fail}}/2\), since each failure erases one of the two parity outcomes. The erasure threshold therefore corresponds to
\begin{equation*}
p_{\mathrm{fail}}^{\mathrm{th}}
=2p_{\mathrm{erasure}}^{\mathrm{th}}
=23.96\%<25\%.
\end{equation*}
Under these assumptions, first-level boosting with \(p_{\mathrm{fail}}=1/4\) is insufficient. We therefore use second-level Ewert-van Loock boosted fusions, with an ideal failure probability of \(1/8\)~\cite{3/4BSM} .
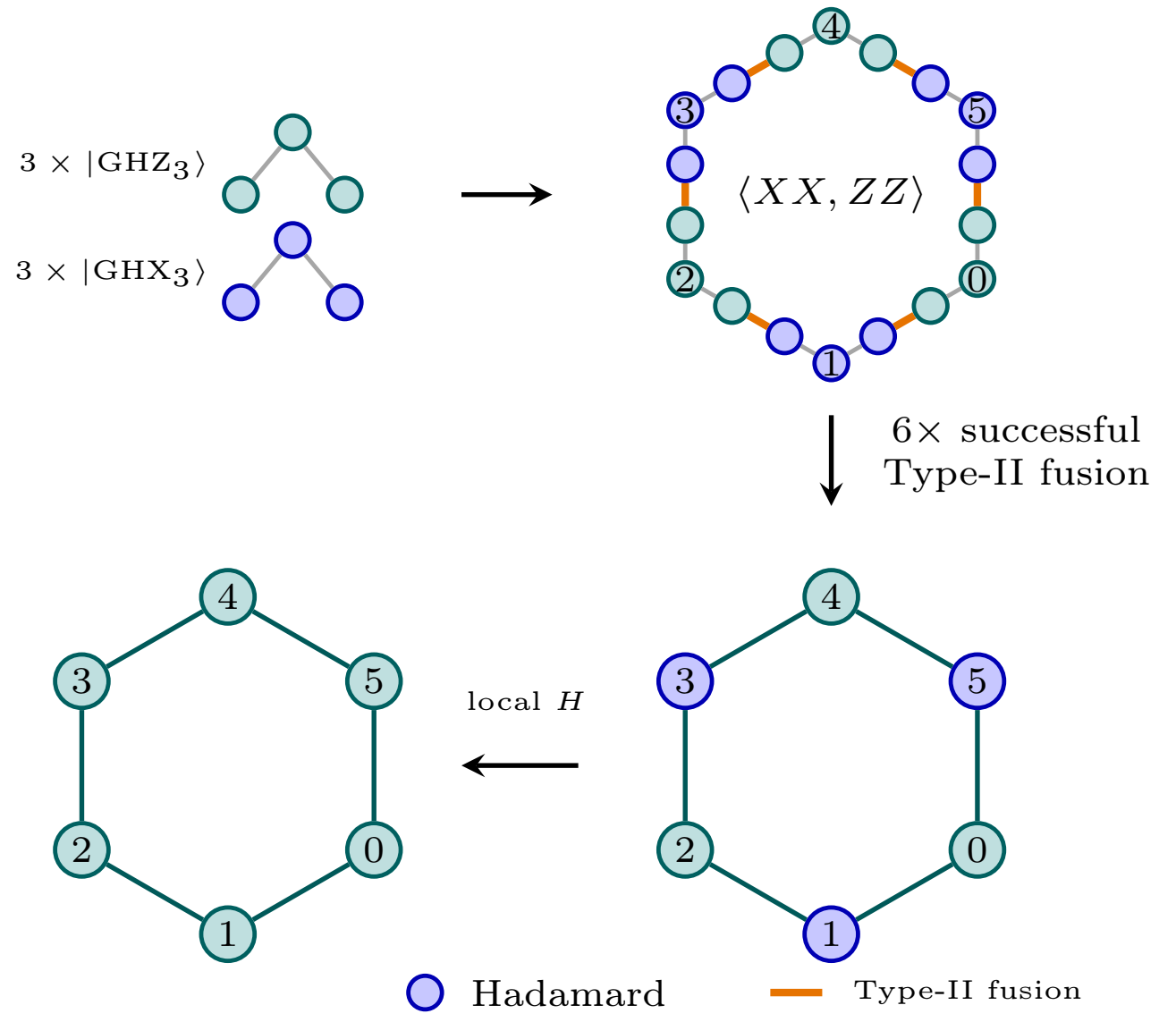
\begin{figure}[ht]
\centering
\resizebox{\columnwidth}{!}{%
\begin{tikzpicture}[
    scale=0.85,
    thick,
    >=stealth,
    ghzphoton/.style={
        circle,
        draw=teal!75!black,
        fill=teal!25,
        minimum size=2.2mm,
        inner sep=0pt,
        font=\scriptsize
    },
    ghxphoton/.style={
        circle,
        draw=blue!70!black,
        fill=blue!22,
        minimum size=2.2mm,
        inner sep=0pt,
        font=\scriptsize
    },
    outphoton/.style={
        circle,
        draw=teal!75!black,
        fill=teal!25,
        minimum size=3.5mm,
        inner sep=0pt,
        font=\scriptsize
    },
    outphotonH/.style={
        circle,
        draw=blue!70!black,
        fill=blue!22,
        minimum size=3.5mm,
        inner sep=0pt,
        font=\scriptsize
    },
    seededge/.style={
        draw=gray!70,
        line width=0.8pt
    },
    ringedge/.style={
        draw=teal!70!black,
        line width=0.9pt
    },
    fusionedge/.style={
        draw=orange!90!black,
        line width=1.4pt
    },
    process/.style={
        ->,
        draw=black,
        line width=0.9pt
    },
    hgate/.style={
        rectangle,
        draw=green!50!black,
        fill=green!15,
        minimum width=3.8mm,
        minimum height=3.8mm,
        inner sep=0pt,
        font=\scriptsize
    },
    textnode/.style={
        rectangle,
        draw=none,
        fill=none,
        align=center,
        font=\scriptsize
    }
]

\def\R{1.3}

% ============================================================
% (a) Three GHZ3 and three GHX3 seed states
% ============================================================
\begin{scope}[shift={(0,2.2)}]

% GHZ3 template
\node[ghzphoton] (gz0) at (0,0.48) {};
\node[ghzphoton] (gz1) at (-0.40,0.00) {};
\node[ghzphoton] (gz2) at (0.40,0.00) {};

\draw[seededge] (gz0)--(gz1);
\draw[seededge] (gz0)--(gz2);

\node[textnode,anchor=east] at (-0.5,0.23)
    {\tiny{\(3 \times \ket{\mathrm{GHZ}_3}\)}};

% GHX3 template
\node[ghxphoton] (gx0) at (0,-0.35) {};
\node[ghxphoton] (gx1) at (-0.40,-0.83) {};
\node[ghxphoton] (gx2) at (0.40,-0.83) {};

\draw[seededge] (gx0)--(gx1);
\draw[seededge] (gx0)--(gx2);

\node[textnode,anchor=east,font=\scriptsize] at (-0.5,-0.60)
    {\tiny{\(3 \times\ket{\mathrm{GHX}_3}\)}};

\end{scope}

% \node[textnode] at (0,3.2)
%     {\textbf{(a)} Photonic seed states};

% Routing arrow
\draw[process] (1.30,2.2) -- (2,2.2);
% \node[textnode] at (2.08,2.55)
%     {routing};

% ============================================================
% (b) Cyclic Type-II fusion network
% ============================================================
\begin{scope}[shift={(4.15,2.2)}]

% Central photons that survive the fusions
\node[ghzphoton] (c0) at (-30:\R) {0};
\node[ghxphoton] (c1) at (-90:\R) {1};
\node[ghzphoton] (c2) at (-150:\R) {2};
\node[ghxphoton] (c3) at (150:\R) {3};
\node[ghzphoton] (c4) at (90:\R) {4};
\node[ghxphoton] (c5) at (30:\R) {5};

% Fusion arms on edge 0--1
\node[ghzphoton] (a01) at ($(c0)!0.32!(c1)$) {};
\node[ghxphoton] (a10) at ($(c0)!0.68!(c1)$) {};

% Edge 1--0
\draw[seededge] (c0)--(a01);
\draw[seededge] (c1)--(a10);

% Edge 1--2
\node[ghxphoton] (a12) at ($(c1)!0.32!(c2)$) {};
\node[ghzphoton] (a21) at ($(c1)!0.68!(c2)$) {};

% Edge 2--3
\node[ghzphoton] (a23) at ($(c2)!0.32!(c3)$) {};
\node[ghxphoton] (a32) at ($(c2)!0.68!(c3)$) {};

% Edge 3--4
\node[ghxphoton] (a34) at ($(c3)!0.32!(c4)$) {};
\node[ghzphoton] (a43) at ($(c3)!0.68!(c4)$) {};

% Edge 4--5
\node[ghzphoton] (a45) at ($(c4)!0.32!(c5)$) {};
\node[ghxphoton] (a54) at ($(c4)!0.68!(c5)$) {};

% Edge 5--0
\node[ghxphoton] (a50) at ($(c5)!0.32!(c0)$) {};
\node[ghzphoton] (a05) at ($(c5)!0.68!(c0)$) {};

\draw[seededge] (c0)--(a05);

\draw[seededge] (c1)--(a12);

\draw[seededge] (c2)--(a21);
\draw[seededge] (c2)--(a23);

\draw[seededge] (c3)--(a32);
\draw[seededge] (c3)--(a34);

\draw[seededge] (c4)--(a43);
\draw[seededge] (c4)--(a45);

\draw[seededge] (c5)--(a54);
\draw[seededge] (c5)--(a50);

% Six Type-II fusion measurements
\draw[fusionedge] (a01)--(a10);
\draw[fusionedge] (a12)--(a21);
\draw[fusionedge] (a23)--(a32);
\draw[fusionedge] (a34)--(a43);
\draw[fusionedge] (a45)--(a54);
\draw[fusionedge] (a50)--(a05);

\node[textnode] at (0,0)
    {\(\langle XX,ZZ\rangle\)};

\end{scope}

% \node[textnode] at (4.15,3.2)
%     {\textbf{(b)} Cyclic fusion arrangement};

% Fusion arrow
\draw[process] (4.15,0.5) -- (4.15,-0.2);
\node[textnode,anchor=west] at (4.40,0.2)
    {\(6\times\) successful\\Type-II fusion};

% ============================================================
% (c) Six surviving photons, before local corrections
% ============================================================
\begin{scope}[shift={(4.15,-2.2)}]

\node[outphoton] (p0) at (-30:\R) {0};
\node[outphotonH] (p1) at (-90:\R) {1};
\node[outphoton] (p2) at (-150:\R) {2};
\node[outphotonH] (p3) at (150:\R) {3};
\node[outphoton] (p4) at (90:\R) {4};
\node[outphotonH] (p5) at (30:\R) {5};

\draw[ringedge] (p0)--(p1);
\draw[ringedge] (p1)--(p2);
\draw[ringedge] (p2)--(p3);
\draw[ringedge] (p3)--(p4);
\draw[ringedge] (p4)--(p5);
\draw[ringedge] (p5)--(p0);

% Local Hadamards required on the GHX-derived outputs
% \node[hgate] at (-90:1.28) {\(H\)};
% \node[hgate] at (150:1.28) {\(H\)};
% \node[hgate] at (30:1.28) {\(H\)};

\end{scope}

% \node[textnode] at (4.15,-3.35)
%     {\textbf{(c)} Post-fusion state};

% Local correction arrow
\draw[process] (2.2,-2.2) -- (1.3,-2.2);
\node[textnode] at (1.8,-1.7)
    {\tiny{local \(H\)}};

% ============================================================
% (d) Canonical photonic six-ring
% ============================================================
\begin{scope}[shift={(-0.5,-2.2)}]

\foreach \name/\lab/\ang in {
    q0/0/-30,
    q1/1/-90,
    q2/2/-150,
    q3/3/150,
    q4/4/90,
    q5/5/30
}{
    \node[outphoton] (\name) at (\ang:\R) {\lab};
}

\draw[ringedge] (q0)--(q1);
\draw[ringedge] (q1)--(q2);
\draw[ringedge] (q2)--(q3);
\draw[ringedge] (q3)--(q4);
\draw[ringedge] (q4)--(q5);
\draw[ringedge] (q5)--(q0);

\end{scope}

% \node[textnode] at (0,-3.35)
%     {\textbf{(d)} Photonic six-ring};

% ============================================================
% Legend
% ============================================================
% \node[ghzphoton] at (0.35,-3.95) {};
% \node[textnode,anchor=west] at (0.56,-3.95)
%     {\(\mathrm{GHZ}_3\) photon};

\node[ghxphoton] at (1.02,-3.95) {};
\node[textnode,anchor=west] at (1.23,-3.95)
    { Hadamard };

\draw[fusionedge] (3.68,-3.95)--(4.08,-3.95);
\node[textnode,anchor=west] at (4.17,-3.95)
    {\tiny{Type-II fusion}};

\end{tikzpicture}%
}

\caption[All-photonic construction of a six-ring resource state]{%
\justifying
All-photonic construction of a six-ring resource state from three
\(\ket{\mathrm{GHZ}_3}\) and three
\(\ket{\mathrm{GHX}_3}=H^{\otimes3}\ket{\mathrm{GHZ}_3}\)
seed states. Neighboring states are connected through successful
Type-II fusion measurements of the \(XX\) and \(ZZ\) parities.
Blue nodes indicate qubits on which a local Hadamardoperation
is required to obtain the canonical six-ring graph state.
}
\label{fig:photonic-six-ring-generation}
\end{figure}

We extend this description by including the distinguishability-induced noise generated during both
resource preparation and fusion measurements, accounting for parity errors and additional erasures. We compare two preparation schemes for the same ideal six-ring graph state: an emitter-assisted construction using six quantum emitters and RUS \(\mathrm{CZ}\) gates, and an all-photonic construction assembled from smaller photonic resource states. The prepared states are then connected using the same fusion network and measurement noise model. This allows us to assess how the resource-generation method affects the distinguishability threshold.

\subsection{All-photonic construction}
%Photonic 6 ring

In the all-photonic construction, the six-ring state is assembled from three \(\ket{\mathrm{GHZ}_3}\) states and three \(\ket{\mathrm{GHX}_3}=H^{\otimes 3}\ket{\mathrm{GHZ}_3}\) states as shown in Fig.~\ref{fig:photonic-six-ring-generation}. The six seed states are prepared in parallel using the generation scheme of Ref.~\cite{Bartolucci2021Creation}. We explicitly account for partial photon distinguishability during seed-state preparation using the noisy postselected \(\mathrm{GHZ}_3\) state derived in Appendix~\ref{app:ghz3-noise}. For this preparation, we use unboosted Type-II fusions. Although boosting would increase the success probability of each fusion, it would require additional ancillary photons and introduce further distinguishability errors. Since the six-ring states are prepared offline, the lower success probability of the unboosted fusions affects only the preparation rate and
resource overhead.
Therefore, the preparation noise model includes distinguishability errors from both the seed states and the successful Type-II fusions used to assemble the ring.
%RUS 6 ring
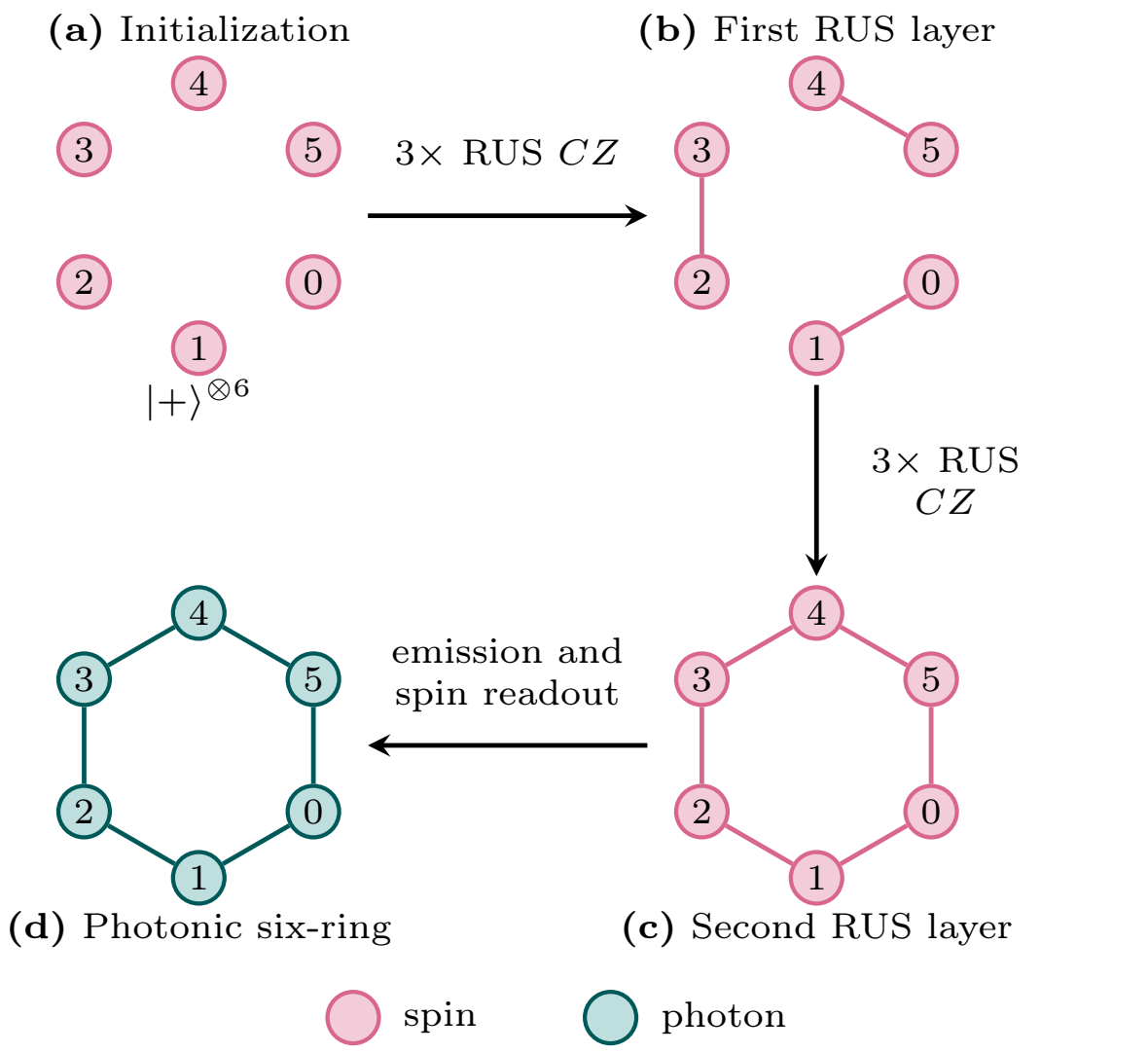
\begin{figure}[!t]
\centering
\begin{tikzpicture}[
    thick,
    >=stealth,
    spin/.style={
        circle,
        draw=purple!60,
        fill=purple!20,
        minimum size=3.5mm,
        inner sep=0pt,
        font=\scriptsize
    },
    photon/.style={
        circle,
        draw=teal!70!black,
        fill=teal!25,
        minimum size=3.5mm,
        inner sep=0pt,
        font=\scriptsize
    },
    spinedge/.style={
        draw=purple!60,
        line width=0.9pt
    },
    photonedge/.style={
        draw=teal!70!black,
        line width=0.9pt
    },
    process/.style={
        ->,
        draw=black,
        line width=0.9pt
    },
    textnode/.style={
        rectangle,
        draw=none,
        fill=none,
        align=center,
        font=\scriptsize
    }
]

\def\radius{0.9}

% ============================================================
% (a) Initialized spin emitters: upper left
% ============================================================
\begin{scope}[shift={(0,1.8)}]
\foreach \name/\lab/\ang in {
    a4/4/90,
    a5/5/30,
    a0/0/-30,
    a1/1/-90,
    a2/2/-150,
    a3/3/150
}{
    \node[spin] (\name) at (\ang:\radius) {\lab};
}

\node[textnode] at (0,-1.25)
    {\(\ket{+}^{\otimes 6}\)};
\end{scope}

\node[textnode] at (0,3.05)
    {\textbf{(a)} Initialization};

% First RUS layer
\draw[process] (1.15,1.8) -- (3.05,1.8);
\node[textnode] at (2.1,2.22)
    {\(3\times\) RUS \(CZ\)};

% ============================================================
% (b) Three spin pairs: upper right
% ============================================================
\begin{scope}[shift={(4.2,1.8)}]
\foreach \name/\lab/\ang in {
    b4/4/90,
    b5/5/30,
    b0/0/-30,
    b1/1/-90,
    b2/2/-150,
    b3/3/150
}{
    \node[spin] (\name) at (\ang:\radius) {\lab};
}

\draw[spinedge] (b0)--(b1);
\draw[spinedge] (b2)--(b3);
\draw[spinedge] (b4)--(b5);
\end{scope}

\node[textnode] at (4.2,3.05)
    {\textbf{(b)} First RUS layer};

% Second RUS layer: downward arrow
\draw[process] (4.2,0.65) -- (4.2,-0.65);
\node[textnode,anchor=west] at (4.45,0)
    {\(3\times\) RUS\\\(CZ\)};

% ============================================================
% (c) Completed spin ring: lower right
% ============================================================
\begin{scope}[shift={(4.2,-1.8)}]
\foreach \name/\lab/\ang in {
    c4/4/90,
    c5/5/30,
    c0/0/-30,
    c1/1/-90,
    c2/2/-150,
    c3/3/150
}{
    \node[spin] (\name) at (\ang:\radius) {\lab};
}

\draw[spinedge] (c0)--(c1);
\draw[spinedge] (c1)--(c2);
\draw[spinedge] (c2)--(c3);
\draw[spinedge] (c3)--(c4);
\draw[spinedge] (c4)--(c5);
\draw[spinedge] (c5)--(c0);
\end{scope}

\node[textnode] at (4.2,-3.05)
    {\textbf{(c)} Second RUS layer};

% Emission: arrow points from lower right to lower left
\draw[process] (3.05,-1.8) -- (1.15,-1.8);
\node[textnode] at (2.1,-1.32)
    {emission and\\spin readout};

% ============================================================
% (d) Photonic six-ring: lower left
% ============================================================
\begin{scope}[shift={(0,-1.8)}]
\foreach \name/\lab/\ang in {
    d4/4/90,
    d5/5/30,
    d0/0/-30,
    d1/1/-90,
    d2/2/-150,
    d3/3/150
}{
    \node[photon] (\name) at (\ang:\radius) {\lab};
}

\draw[photonedge] (d0)--(d1);
\draw[photonedge] (d1)--(d2);
\draw[photonedge] (d2)--(d3);
\draw[photonedge] (d3)--(d4);
\draw[photonedge] (d4)--(d5);
\draw[photonedge] (d5)--(d0);
\end{scope}

\node[textnode] at (0,-3.05)
    {\textbf{(d)} Photonic six-ring};

% Legend
\node[spin] at (1.05,-3.65) {};
\node[textnode,anchor=west] at (1.27,-3.65) {spin};

\node[photon] at (2.80,-3.65) {};
\node[textnode,anchor=west] at (3.02,-3.65) {photon};

\end{tikzpicture}
\caption{Emitter-assisted generation of a photonic six-ring resource state~\cite{SPOQC,Lim2005}. Two successive layers of parallel RUS \(CZ\) operations prepare the six-ring graph state on the spin emitters. Photon emission and spin readout then transfer the graph state to the
photonic qubits.}
\label{fig:rus-six-ring-generation}
\end{figure}

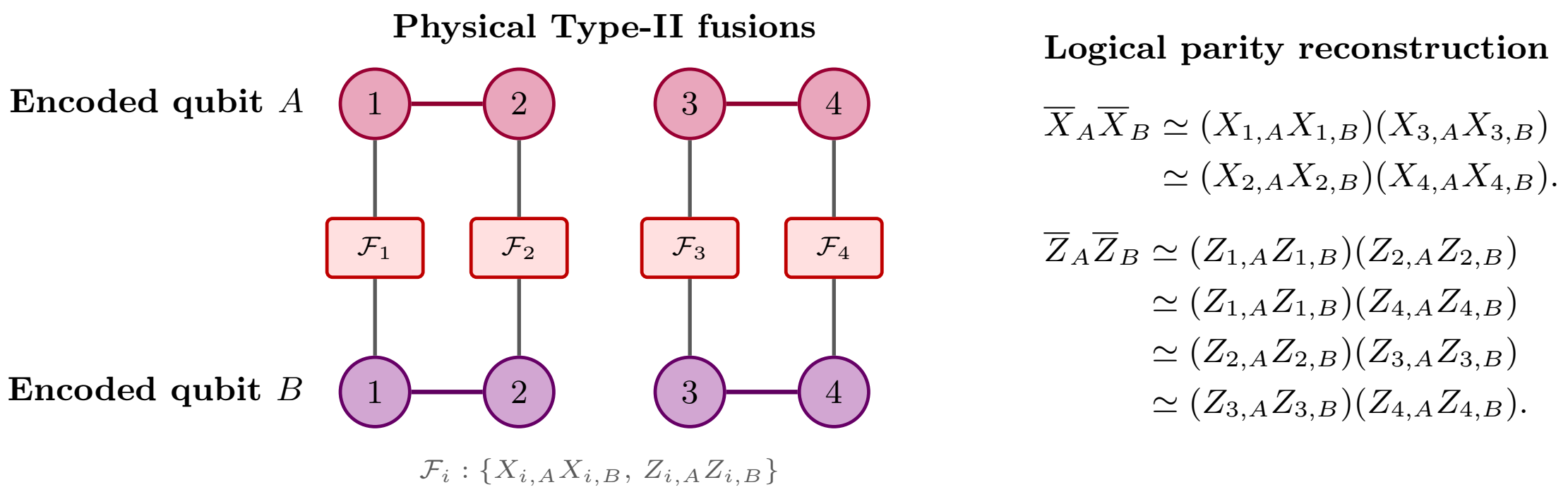
\begin{figure*}[t!]
    \centering
    \begin{tikzpicture}[
        x=1cm,
        y=1cm,
        qubitA/.style={
            circle,
            draw=purple!80!black,
            fill=purple!35,
            minimum size=6.5mm,
            inner sep=0pt,
            line width=0.9pt,
            font=\small
        },
        qubitB/.style={
            circle,
            draw=violet!80!black,
            fill=violet!35,
            minimum size=6.5mm,
            inner sep=0pt,
            line width=0.9pt,
            font=\small
        },
        graphA/.style={
            draw=purple!75!black,
            line width=1.2pt
        },
        graphB/.style={
            draw=violet!75!black,
            line width=1.2pt
        },
        fusion/.style={
            rectangle,
            rounded corners=1.5pt,
            draw=red!75!black,
            fill=red!12,
            minimum width=9mm,
            minimum height=5.5mm,
            line width=0.9pt,
            font=\scriptsize
        },
        fusionline/.style={
            draw=gray!70!black,
            line width=1pt
        }
    ]

    %------------------------------------------------
    % Encoded qubit A
    %------------------------------------------------
    \node[draw=none, anchor=east, font=\bfseries] at (-0.55,1.35)
        {Encoded qubit \(A\)};

    \node[qubitA] (A1) at (0,1.35)   {\(1\)};
    \node[qubitA] (A2) at (1.35,1.35){\(2\)};
    \node[qubitA] (A3) at (2.95,1.35){\(3\)};
    \node[qubitA] (A4) at (4.30,1.35){\(4\)};

    % Internal graph edges: 1--2 and 3--4
    \draw[graphA] (A1) -- (A2);
    \draw[graphA] (A3) -- (A4);

    %------------------------------------------------
    % Encoded qubit B
    %------------------------------------------------
    \node[draw=none, anchor=east, font=\bfseries] at (-0.55,-1.35)
        {Encoded qubit \(B\)};

    \node[qubitB] (B1) at (0,-1.35)   {\(1\)};
    \node[qubitB] (B2) at (1.35,-1.35){\(2\)};
    \node[qubitB] (B3) at (2.95,-1.35){\(3\)};
    \node[qubitB] (B4) at (4.30,-1.35){\(4\)};

    % Internal graph edges: 1--2 and 3--4
    \draw[graphB] (B1) -- (B2);
    \draw[graphB] (B3) -- (B4);

    %------------------------------------------------
    % Transversal physical fusions
    %------------------------------------------------
    \node[draw=none, font=\small\bfseries] at (2.15,2.05)
        {Physical Type-II fusions};

    \node[fusion] (F1) at (0,0)    {\(\mathcal F_1\)};
    \node[fusion] (F2) at (1.35,0) {\(\mathcal F_2\)};
    \node[fusion] (F3) at (2.95,0) {\(\mathcal F_3\)};
    \node[fusion] (F4) at (4.30,0) {\(\mathcal F_4\)};

    \draw[fusionline] (A1) -- (F1);
    \draw[fusionline] (F1) -- (B1);

    \draw[fusionline] (A2) -- (F2);
    \draw[fusionline] (F2) -- (B2);

    \draw[fusionline] (A3) -- (F3);
    \draw[fusionline] (F3) -- (B3);

    \draw[fusionline] (A4) -- (F4);
    \draw[fusionline] (F4) -- (B4);

    % Measurement performed by each fusion
    \node[
        draw=none,
        align=center,
        font=\scriptsize,
        text=gray!70!black
    ] at (2.15,-2.10) {
        \(\mathcal F_i:
        \{X_{i,A}X_{i,B},\,Z_{i,A}Z_{i,B}\}\)
    };

    %------------------------------------------------
    % Arrow toward logical reconstruction
    %------------------------------------------------
    % \draw[
    %     -{Latex[length=2.5mm]},
    %     gray!70!black,
    %     line width=0.9pt
    % ] (4.90,0) -- (5.90,0)
    % node[midway, above=2mm, draw=none, font=\scriptsize]
    %     {reconstruction};

    %------------------------------------------------
    % Logical parity reconstruction — no box
    %------------------------------------------------
    \node[
        draw=none,
        anchor=west,
        font=\bfseries
    ] at (6.15,1.85)
        {Logical parity reconstruction};

    \node[
        draw=none,
        anchor=west,
        align=left,
        font=\small
    ] at (6.15,0.90) {
        $\displaystyle
        \begin{aligned}
        \overline{X}_A\overline{X}_B
        &\simeq
        (X_{1,A}X_{1,B})(X_{3,A}X_{3,B})\\
        &\simeq
        (X_{2,A}X_{2,B})(X_{4,A}X_{4,B}) .
        \end{aligned}$
    };

    \node[
        draw=none,
        anchor=west,
        align=left,
        font=\small
    ] at (6.15,-0.75) {
        $\displaystyle
        \begin{aligned}
        \overline{Z}_A\overline{Z}_B
        &\simeq
        (Z_{1,A}Z_{1,B})(Z_{2,A}Z_{2,B})\\
        &\simeq
        (Z_{1,A}Z_{1,B})(Z_{4,A}Z_{4,B})\\
        &\simeq
        (Z_{2,A}Z_{2,B})(Z_{3,A}Z_{3,B})\\
        &\simeq
        (Z_{3,A}Z_{3,B})(Z_{4,A}Z_{4,B}) .
        \end{aligned}$
    };

    \end{tikzpicture}
    \captionsetup{justification=raggedright, singlelinecheck=false}
    \caption{Transversal Type-II fusion between two encoded qubits using a Z-biased Shor$(2,2)$ code. Each physical fusion \(\mathcal F_i\) measures the \(XX\) and \(ZZ\) parities of the corresponding pair of physical qubits. Equivalent products of these outcomes reconstruct the logical parities \(\overline{X}_A\overline{X}_B\) and \(\overline{Z}_A\overline{Z}_B\).}
    \label{fig:physical-fusions}
\end{figure*}

\subsection{Emitter-based construction}
The emitter-based construction relies on the RUS \(\mathrm{CZ}\) gate introduced in Ref.~\cite{SPOQC}. Six spins are initialized in \(\ket{+}^{\otimes 6}\), and the six edges of the ring are generated using two successive layers of three parallel RUS \(\mathrm{CZ}\) gates. Photon emission subsequently entangles each spin with a photonic qubit and spin readout transfers the ring graph state to the six emitted photons, up to known Pauli-frame corrections. The preparation procedure is shown in Fig.~\ref{fig:rus-six-ring-generation}.

Partial distinguishability in a successful RUS \(\mathrm{CZ}\) operation is described by the correlated \(\mathrm{Z_1Z_2}\) channel introduced in
Sec.~\ref{sec:rus-noise}.

\section{Shor(2,2) encoded six-ring}
\label{sec:shor_22_encoded_six_ring}

Encoding a photonic resource state can improve its tolerance to photon loss by distributing the logical information redundantly across several photons. This improvement comes at the cost of additional photons and physical fusion
measurements. Here, we consider the six-ring resource state encoded using the Shor(2,2) code, also known as the quantum parity code. This $\llbracket4,1,2\rrbracket$ code encodes each logical qubit into four physical photons.
Under the linear-optical loss model of Ref.~\cite{FBQC}, the encoded fusion network achieves a per-photon loss threshold of $p_{\mathrm{loss}}^{\star}=2.7\%$ using boosted physical fusions with $p_{\mathrm{fail}}=1/4$. For comparison, the unencoded six-ring network achieves \(p_{\mathrm{loss}}^\star=0.75\%\), but requires more strongly boosted fusions with \(p_{\mathrm{fail}}=1/8\).

\subsection{Shor(2,2) code}
Using the stabilizer formalism \cite{Gottesman1997}, we can describe an encoded qubit $i$ using the Shor(2,2) code through the stabilizer group
\begin{equation}
\mathcal{S}_i
=
\left\langle
X_{i,1}X_{i,2}X_{i,3}X_{i,4},
Z_{i,1}Z_{i,3},
Z_{i,2}Z_{i,4}
\right\rangle ,
\end{equation}
with logical operators
\begin{equation}
\overline{X}_i=X_{i,1}X_{i,3},
\qquad
\overline{Z}_i=Z_{i,1}Z_{i,2}.
\end{equation}
This \(Z\)-biased convention corresponds to an $X$-repetition code concatenated on top of a $Z$-repetition code.
The complementary \(X\)-biased convention reverses this concatenation order, giving
\begin{equation}
\mathcal{S}_i^{(X)}
=
\left\langle
Z_{i,1}Z_{i,2}Z_{i,3}Z_{i,4},
X_{i,1}X_{i,3},
X_{i,2}X_{i,4}
\right\rangle ,
\end{equation}
with logical operators chosen as
\begin{equation}
\overline{X}_i=X_{i,1}X_{i,2},
\qquad
\overline{Z}_i=Z_{i,1}Z_{i,3}.
\end{equation}

A logical fusion between encoded qubits \(A\) and \(B\) is implemented transversally using four physical Type-II fusions, as illustrated in Fig.~\ref{fig:physical-fusions} for the Z-biased encoding. The redundancy among the logical operator representatives allows the logical parities \(\overline{X}_A\overline{X}_B\) and \(\overline{Z}_A\overline{Z}_B\) to be reconstructed from several combinations
of the physical outcomes. This redundancy is responsible for the enhanced tolerance to heralded erasures.

\subsection{Encoded resource-state generation}
\label{sec:encoded-resource-generation}
Wein et al.~\cite{Wein2025} proposed three architectures for generating the Shor(2,2) encoded six-ring resource state: an all-photonic construction based on photonic seed states and fusion measurements, a construction based on emitter-generated caterpillar graph states, and a hybrid construction in which spin emitters are entangled using repeat-until-success (RUS) gates. In this work, we restrict our analysis to the all-photonic and RUS constructions. These schemes represent two limiting resource trade-offs: the all-photonic construction requires multiple sources, fusion stages, and spatial multiplexing, whereas the RUS construction reduces the optical and multiplexing overhead by reusing stationary emitter qubits over several time steps.

\subsubsection{All-photonic construction}

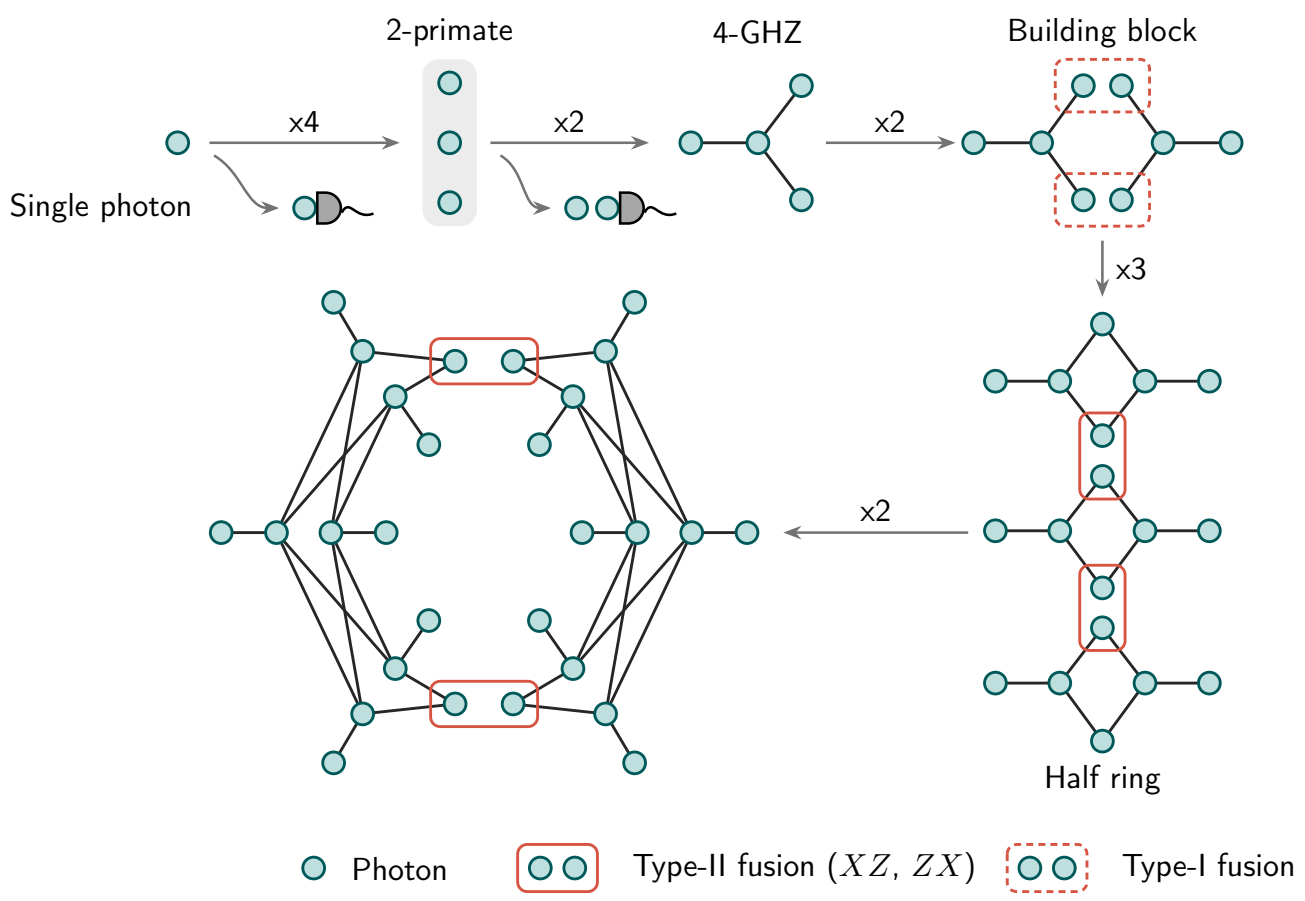
\begin{figure}[!t]
    \centering

\scalebox{0.6}{
    \definecolor{arrowgray}{gray}{0.45}
\definecolor{boxred}{RGB}{214,84,66}

\tikzset{
  % --- one-line swap: replace by  fill=boxred!85, draw=black  for red photons
  phot/.style={circle, draw=teal!70!black, fill=teal!25, inner sep=0pt,
               minimum size=2.4mm, line width=0.9pt},
  gedge/.style={draw=black!85, line width=0.9pt},
  fbox/.style={draw=boxred, line width=0.9pt, rounded corners=3pt},   % Type-II fusion
  dbox/.style={draw=boxred, dash pattern=on 2.2pt off 1.6pt, line width=0.9pt,
               rounded corners=3pt},
  garrow/.style={-{Stealth[length=2mm]}, draw=arrowgray, line width=0.8pt},
  alabel/.style={font=\sffamily\normalsize, text=black}
}

% photon detector symbol, left-centre at (#1,#2)
\newcommand{\detector}[2]{%
  \begin{scope}[shift={(#1,#2)}]
    \path[fill=gray!70, draw=black, line width=0.8pt]
      (0,-17)--(9,-17) arc[start angle=-90,end angle=90,radius=17]--(0,17)--cycle;
    \draw[black, line width=0.8pt] (27,5) .. controls (36,-12) and (42,14) .. (62,6);
  \end{scope}}

\begin{tikzpicture}[x=0.01cm, y=-0.01cm]

%% ===================== ROW 1 =====================

% ---- single photon (+ measured photon)
\node[phot] at (0,0) {};
\node[alabel, anchor=north east] at (28,44) {Single photon};
\draw[garrow] (35,0)--(245,0);          \node[alabel] at (140,-24) {x4};
\draw[garrow] (40,14) to[out=-30,in=180] (112,72);
\node[phot] at (140,72) {};  \detector{154}{72}

% ---- 2-primate
\fill[gray!15, rounded corners=5pt] (270,-92) rectangle (330,92);
\foreach \y in {-66,0,66} \node[phot] at (300,\y) {};
\node[alabel] at (300,-122) {2-primate};
\draw[garrow] (345,0)--(520,0);         \node[alabel] at (432,-24) {x2};
\draw[garrow] (356,14) to[out=-30,in=180] (412,72);
\node[phot] at (440,72) {};  \node[phot] at (474,72) {};  \detector{488}{72}

% ---- 4-GHZ
\draw[gedge] (566,0)--(640,0)--(688,-63);
\draw[gedge] (640,0)--(688,63);
\foreach \p in {(566,0),(640,0),(688,-63),(688,63)} \node[phot] at \p {};
\node[alabel] at (640,-122) {4-GHZ};
\draw[garrow] (715,0)--(858,0);         \node[alabel] at (786,-24) {x2};

% ---- building block (centre x = 1020)
\begin{scope}[shift={(1020,0)}]
  \draw[gedge] (-142,0)--(-67,0)--(-21,-63);
  \draw[gedge] (-67,0)--(-21,63);
  \draw[gedge] (142,0)--(67,0)--(21,-63);
  \draw[gedge] (67,0)--(21,63);
  \foreach \p in {(-142,0),(-67,0),(-21,-63),(21,-63),(-21,63),(21,63),(67,0),(142,0)}
    \node[phot] at \p {};
  \draw[dbox] (-52,-92) rectangle (52,-34);
  \draw[dbox] (-52,34) rectangle (52,92);
  \node[alabel] at (0,-122) {Building block};
\end{scope}

%% ===================== wrap-down arrow (x3) =====================
\draw[garrow] (1020,108)--(1020,170);   \node[alabel] at (1052,139) {x3};

%% ===================== ROW 2 =====================

% ---- half ring: vertical chain of three diamonds, centre (1020,430)
%      side vertices carry a leaf; the two photons facing each other
%      between consecutive diamonds are joined by a Type-II fusion.
\begin{scope}[shift={(1020,430)}]
  % diamond 1
  \draw[gedge] (0,-230)--(-47,-167)--(0,-107)--(47,-167)--cycle;
  \draw[gedge] (-47,-167)--(-118,-167);  \draw[gedge] (47,-167)--(118,-167);
  % diamond 2
  \draw[gedge] (0,-62)--(-47,-2)--(0,61)--(47,-2)--cycle;
  \draw[gedge] (-47,-2)--(-118,-2);      \draw[gedge] (47,-2)--(118,-2);
  % diamond 3
  \draw[gedge] (0,105)--(-47,166)--(0,230)--(47,166)--cycle;
  \draw[gedge] (-47,166)--(-118,166);    \draw[gedge] (47,166)--(118,166);
  \foreach \p in {(0,-230),(-47,-167),(47,-167),(-118,-167),(118,-167),(0,-107),
                  (0,-62),(-47,-2),(47,-2),(-118,-2),(118,-2),(0,61),
                  (0,105),(-47,166),(47,166),(-118,166),(118,166),(0,230)}
    \node[phot] at \p {};
  % Type-II fusions
  \draw[fbox] (-25,-132) rectangle (25,-37);
  \draw[fbox] (-25,36)   rectangle (25,130);
  \node[alabel] at (0,272) {Half ring};
\end{scope}

\draw[garrow] (872,430)--(668,430);     \node[alabel] at (770,406) {x2};

% ---- final 24-photon ring, centre (338,430)
\begin{scope}[shift={(338,430)}]
  \foreach \sx in {-1,1}{
    \coordinate (V\sx) at (\sx*229,0);  \coordinate (I\sx) at (\sx*169,0);
    \foreach \sy in {-1,1}{
      \coordinate (P\sx\sy) at (\sx*134,\sy*200);
      \coordinate (Q\sx\sy) at (\sx*98,\sy*150);
      \coordinate (A\sx\sy) at (\sx*32,\sy*189);
      \coordinate (U\sx\sy) at (\sx*166,\sy*254);
      \coordinate (W\sx\sy) at (\sx*61,\sy*97);
      \draw[gedge] (P\sx\sy)--(A\sx\sy);  \draw[gedge] (Q\sx\sy)--(A\sx\sy);
      \draw[gedge] (P\sx\sy)--(V\sx);     \draw[gedge] (P\sx\sy)--(I\sx);
      \draw[gedge] (Q\sx\sy)--(V\sx);     \draw[gedge] (Q\sx\sy)--(I\sx);
      \draw[gedge] (P\sx\sy)--(U\sx\sy);  \draw[gedge] (Q\sx\sy)--(W\sx\sy);
    }
    \draw[gedge] (V\sx)--(\sx*290,0);     \draw[gedge] (I\sx)--(\sx*108,0);
  }
  \foreach \sx in {-1,1}{
    \node[phot] at (V\sx) {};  \node[phot] at (I\sx) {};
    \node[phot] at (\sx*290,0) {};  \node[phot] at (\sx*108,0) {};
    \foreach \sy in {-1,1}{
      \foreach \n in {P,Q,A,U,W} \node[phot] at (\n\sx\sy) {};
    }
  }
  % Type-II fusions
  \draw[fbox] (-59,-214) rectangle (59,-164);
  \draw[fbox] (-59,164) rectangle (59,214);
\end{scope}

%% ===================== LEGEND =====================
\begin{scope}[shift={(0,800)}]
  \node[phot] at (150,0) {};
  \node[alabel, anchor=west] at (180,0) {Photon};
  \node[phot] at (400,0) {};  \node[phot] at (438,0) {};
  \draw[fbox] (375,-26) rectangle (463,26);
  \node[alabel, anchor=west] at (490,0) {Type-II fusion ($XZ$, $ZX$)};
    \node[phot] at (940,0) {};  \node[phot] at (978,0) {};
  \draw[dbox] (915,-26) rectangle (1003,26);
  \node[alabel, anchor=west] at (1030,0) {Type-I fusion};
\end{scope}

\end{tikzpicture}
}
\caption{All-photonic generation of the Shor$(2,2)$-encoded 6-ring resource state. Four-photon GHZ states are assembled into building blocks using Type-I fusions (dashed boxes), while Type-II fusions (solid boxes) connect the building blocks into half-rings and, subsequently, the final 24-photon state.}
\label{fig:all_photonic}
\end{figure}

The all-photonic construction starts with twelve four-photon GHZ seed states, \(\ket{\mathrm{GHZ}_4}\), prepared using the linear-optical protocol of Ref.~\cite{Bartolucci2021Creation}. These states are combined through twelve Type-I and six Type-II fusions to produce the final 24-photon encoded six-ring resource state, as illustrated in Fig.~\ref{fig:all_photonic}. The distinguishability-induced noise model for the \(\ket{\mathrm{GHZ}_4}\) seed states is derived in Appendix~\ref{app:ghz4-noise}.

Following Ref.~\cite{Wein2025}, both fusion types are assumed to be boosted, with an ideal success probability of 3/4. This choice reduces the preparation overhead and allows our results to be compared directly with the resource estimates of Ref.~\cite{Wein2025}. It also makes use of the boosting already required by the fusion network considered here. Since resource states are prepared offline and fusion failures are heralded, our simulations are conditioned on the success of all preparation fusions.

For each successful boosted Type-II fusion, we apply the distinguishability-induced error channel derived in Appendix~\ref{app:boosted-fusion-noise}. For successful boosted Type-I fusions, we approximate the conditional error channel using the unboosted Type-I channel presented in Sec.~\ref{sec:unboosted}. This approximation neglects possible additional contributions from the ancillary photons used for boosting; deriving the complete distinguishability-induced channel for boosted Type-I fusion is left for future work.

Unboosted fusions could instead be used during resource-state preparation, while retaining boosted fusions in the fusion network. Avoiding the ancillary photons required for boosting could reduce preparation errors and thereby improve the distinguishability threshold. However, for the construction considered here, the ideal average photon consumption increases from approximately 21,170 to 237,327 photons per resource state~\cite{Wein2025}, a factor of approximately 11.2. The choice of preparation fusions therefore presents a potential trade-off between distinguishability tolerance and resource cost.

% \subsubsection{All photonic scheme}

\subsubsection{Emitter-based RUS construction}

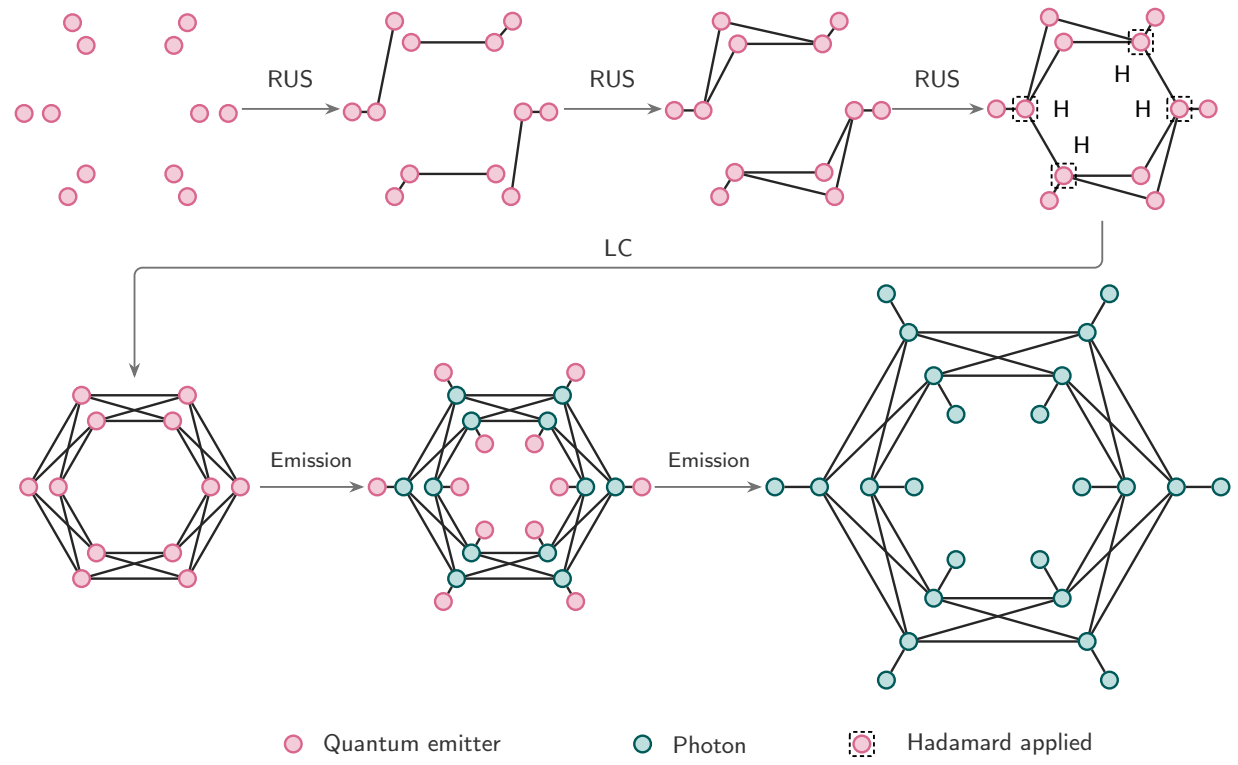
\begin{figure}[!t]

\centering
\definecolor{arrowgray}{gray}{0.45}

\tikzset{
  emit/.style={circle, draw=purple!60, fill=purple!20, inner sep=0pt,
               minimum size=2.2mm, line width=0.9pt},
  phot/.style={emit, draw=teal!70!black, fill=teal!25},
  gedge/.style={draw=black!85, line width=0.9pt},
  hbox/.style={draw=black, dash pattern=on 1.6pt off 1.2pt, line width=0.7pt,
               rounded corners=1.2pt},
  garrow/.style={-{Stealth[length=2mm]}, draw=arrowgray, line width=0.7pt},
  alabel/.style={font=\sffamily\small, text=black!80}
}

% Six-ring core. Args: prefix, cx, cy, R, outer style, inner style
\newcommand{\sixcore}[6]{%
  \foreach \k in {0,...,5}{
    \coordinate (#1co\k) at ({#2+#4*cos(60*\k)},{#3+#4*sin(60*\k)});
    \coordinate (#1ci\k) at ({#2+0.72*#4*cos(60*\k)},{#3+0.72*#4*sin(60*\k)});
  }
  \foreach \k in {0,...,5}{
    \pgfmathtruncatemacro{\n}{mod(\k+1,6)}
    \draw[gedge] (#1co\k)--(#1co\n);
    \draw[gedge] (#1ci\k)--(#1ci\n);
    % \draw[gedge] (#1co\k)--(#1ci\k);
    \draw[gedge] (#1co\k)--(#1ci\n);
    \draw[gedge] (#1co\n)--(#1ci\k);
  }
  \foreach \k in {0,...,5}{
    \node[#5] (#1o\k) at (#1co\k) {};
    \node[#6] (#1i\k) at (#1ci\k) {};
  }
}

% Leaves. Args: prefix, cx, cy, R, outer-leaf style, inner-leaf style
\newcommand{\sixleaves}[6]{%
  \foreach \k in {0,...,5}{
    \node[#5] (#1lo\k) at ({#2+1.25*#4*cos(60*\k)},{#3+1.25*#4*sin(60*\k)}) {};
    \node[#6] (#1li\k) at ({#2+0.47*#4*cos(60*\k)},{#3+0.47*#4*sin(60*\k)}) {};
    \draw[gedge] (#1o\k)--(#1lo\k);
    \draw[gedge] (#1i\k)--(#1li\k);
  }
}
% Pixel-like coordinates (y pointing down)
\scalebox{0.5}{
\begin{tikzpicture}[x=0.02cm, y=-0.02cm]

% =====================  ROW 1 : emitter graph construction  =====================

% ---------------- Panel 1: 6 pairs of emitters ----------------
\begin{scope}[shift={(-96,-490)}]
\foreach \x/\y in {130/433, 139/448,  206/433, 197/448,
                   98/493, 116/493,   214/493, 233/493,
                   139/533, 127/548,  197/533, 206/548}
  \node[emit] at (\x,\y) {};
\end{scope}

% ---------------- Panel 2: after first RUS ----------------
\begin{scope}[shift={(115,-296)}]
\coordinate (a1) at (132,238); \coordinate (a2) at (143,252);
\coordinate (a3) at (198,252); \coordinate (a4) at (210,238);
\coordinate (a5) at (104,298); \coordinate (a6) at (120,298);
\draw[gedge] (a5)--(a6)--(a1);
\draw[gedge] (a2)--(a3)--(a4);
\coordinate (c1) at (142,339); \coordinate (c2) at (199,339);
\coordinate (c3) at (132,354); \coordinate (c4) at (209,354);
\coordinate (c5) at (217,298); \coordinate (c6) at (234,298);
\draw[gedge] (c3)--(c1)--(c2);
\draw[gedge] (c4)--(c5)--(c6);
\foreach \n in {a1,a2,a3,a4,a5,a6,c1,c2,c3,c4,c5,c6}
  \node[emit] at (\n) {};
\end{scope}

% ---------------- Panel 3: after second RUS ----------------
\begin{scope}[shift={(180,-160)}]
\coordinate (p1) at (283,102); \coordinate (p2) at (294,117);
\coordinate (p3) at (350,117); \coordinate (p4) at (361,102);
\coordinate (p5) at (252,161); \coordinate (p6) at (271,161);
\draw[gedge] (p1)--(p3);
\draw[gedge] (p2)--(p3);
\draw[gedge] (p3)--(p4);
\draw[gedge] (p1)--(p6);
\draw[gedge] (p2)--(p6);
\draw[gedge] (p5)--(p6);
\coordinate (q1) at (371,161); \coordinate (q2) at (389,161);
\coordinate (q3) at (292,202); \coordinate (q4) at (282,218);
\coordinate (q5) at (351,202); \coordinate (q6) at (358,218);
\draw[gedge] (q1)--(q2);
\draw[gedge] (q1)--(q5);
\draw[gedge] (q1)--(q6);
\draw[gedge] (q6)--(q3);
\draw[gedge] (q3)--(q5);
\draw[gedge] (q3)--(q4);
\foreach \n in {p1,p2,p3,p4,p5,p6,q1,q2,q3,q4,q5,q6}
  \node[emit] at (\n) {};
\end{scope}

% ---------------- Panel 4: ring with Hadamards ----------------
\begin{scope}[shift={(184,-161)}]
\foreach \k in {0,...,5}{
  \coordinate (hc\k) at ({531+51*cos(60*\k)},{161+51*sin(60*\k)});
  \coordinate (hl\k) at ({531+70*cos(60*\k)},{161+70*sin(60*\k)});
}
\foreach \k in {0,...,5}{
  \pgfmathtruncatemacro{\n}{mod(\k+1,6)}
  \draw[gedge] (hc\k)--(hc\n);
}
\draw[gedge] (hc0)--(hl0);
\draw[gedge] (hc2)--(hl2);
\draw[gedge] (hc3)--(hl3);
\draw[gedge] (hc5)--(hl5);
\draw[gedge] (hl4)--(hc3); \draw[gedge] (hl4)--(hc5);
\draw[gedge] (hl1)--(hc0); \draw[gedge] (hl1)--(hc2);
\foreach \k in {0,...,5}{ \node[emit] at (hc\k) {}; \node[emit] at (hl\k) {}; }
\foreach \k in {0,2,3,5}{
  \draw[hbox] (hc\k) ++(-8,-8) rectangle ++(16,16);
  \node[font=\sffamily\small] at ({531+27*cos(60*\k)},{161+27*sin(60*\k)}) {H};
}
\end{scope}

% =====================  ROW 2 : conversion to photons  =====================

% ---------------- Panel 5: after local complementation ----------------
\begin{scope}[shift={(-615,-46)}]
\sixcore{lc}{690}{296}{70}{emit}{emit}
\end{scope}

% ---------------- Panel 6: emission in progress ----------------
\begin{scope}[shift={(-365,-240)}]
\sixcore{em}{688}{490}{70}{phot}{phot}
\sixleaves{em}{688}{490}{70}{emit}{emit}
\end{scope}

% ---------------- Panel 7: final 24-photon state ----------------
\begin{scope}[shift={(218,-142)}]
\sixcore{ph}{428}{392}{118}{phot}{phot}
\sixleaves{ph}{428}{392}{118}{phot}{phot}
\end{scope}

% =====================  Arrows  =====================
\draw[garrow] (146,0)--(211,0);   \node[alabel] at (178,-20) {RUS};
\draw[garrow] (359,0)--(424,0);   \node[alabel] at (391,-20) {RUS};
\draw[garrow] (576,0)--(637,0);   \node[alabel] at (606,-20) {RUS};

% row wrap: P4 -> P5
\draw[garrow, rounded corners=3pt] (715,74)--(715,105)--(75,105)--(75,178);
\node[alabel] at (395,92) {LC};

\draw[garrow] (158,250)--(227,250);  \node[alabel, font=\sffamily\footnotesize] at (192,232) {Emission};
\draw[garrow] (419,250)--(490,250);  \node[alabel, font=\sffamily\footnotesize] at (455,232) {Emission};

% =====================  Legend  =====================
\node[emit] at (180,420) {};
\node[alabel, anchor=west] at (194,420) {Quantum emitter};
\node[phot] at (411,420) {};
\node[alabel, anchor=west] at (425,420) {Photon};
\node[emit] at (556,420) {};
\draw[hbox] (556,420) ++(-8,-8) rectangle ++(16,16);
\node[alabel, anchor=west] at (580,420) {Hadamard applied};

\end{tikzpicture}
}

\caption{
Generation of the Shor$(2,2)$-encoded six-ring resource state from quantum emitters using repeat-until-success (RUS) gates. Purple and green vertices represent quantum emitters and photons, respectively. Dashed boxes mark emitters on which a Hadamard gate is applied. Starting from twelve emitters, successive RUS CZ operations, Hadamard gates and a local complementation (LC) build the emitter graph. Two emission cycles, each followed by emitter measurements transfer the graph onto the final 24-photon graph state.}
\label{fig:rus-generation}
\end{figure}

The emitter-based construction uses twelve quantum emitters to generate the resource state sequentially. The emitters are first connected through RUS \(\mathrm{CZ}\) operations. Local Hadamard gates are then applied to the four emitters highlighted in Fig.~\ref{fig:rus-generation}, transforming the emitter graph into the required central structure. Two successive emission cycles generate two photonic qubits from each emitter. Finally, the emitters are measured and disentangled, leaving the 24-photon Shor\((2,2)\)-encoded six-ring resource state.

We model each successful RUS \(\mathrm{CZ}\) operation using the \(Z_1Z_2\) distinguishability channel introduced in Sec.~\ref{sec:rus-noise} and derived in Ref.~\cite{SPOQC}. RUS failures are heralded and the operation is repeated until success. Local Hadamard gates, photon emission and emitter measurements are assumed ideal.

\section{Hardware-motivated FBQC simulation}
\label{sec : Hardware-motivated FBQC simulation}
\label{sec : Hardware-motivated FBQC simulation}

Photon-loss tolerance in fusion-based quantum computation has been extensively studied across a range of schemes~\cite{FBQC,
bartolucci2025comparison,bombin2023fault,bombin2023increasing,
lobl2026subthreshold,litinski2025blocklet,PaesaniBrown2023,
bell2023optimizing}. Encoding resource states, for example by applying the Shor$(2,2)$ code to each qubit of a six-ring, can improve loss tolerance. However, preparing larger resource states requires additional operations that can introduce further physical errors. The impact of these errors is often assessed using abstract Pauli noise models, which does not capture the structure of hardware-specific noise.

In this section, we estimate distinguishability thresholds for fusion networks built from unencoded and Shor$(2,2)$-encoded six-ring resource states, using either all-photonic or emitter-assisted RUS preparation schemes. Our circuit-level simulations incorporate the distinguishability-induced errors in both resource-state
preparation and fusion measurements, explicitly resolving the physical operations of the Shor$(2,2)$ encoding. Although thresholds alone do not establish
scalability~\cite{lobl2026subthreshold}, they provide a benchmark for comparing these constructions and assessing their requirements against current hardware capabilities.

\subsection{Threshold simulation}

To accurately simulate a fusion network with hardware realistic noise we proceed as follows:  we build the fusion network as a noiseless \emph{Stim} circuit \cite{gidney2021stim}; we sample heralded noise and failures before including errors in the \emph{Stim} circuit; we sample and decode the circuit from its \emph{detector error model}.

All physics enters through the analytically computed noise models: the resource-state error channels and the fusion error probabilities.
The  simulation only samples from these error models and does not simulate linear-optics.

\subsubsection{Network Construction} 
We simulate the fusion networks at the circuit level, including the Shor-(2,2) encoded one. The construction of these circuits involves several choices, described below, each made to reproduce the original FBQC scheme faithfully.

For a code distance $d$, resource states are arranged on the
sites of a three-dimensional lattice. Each resource state
contains either 6 physical qubits for an unencoded six-ring
or 24 physical qubits for a Shor$(2,2)$-encoded six-ring. In an encoded six-ring, each Shor(2,2) code is randomly biased towards \(X\) or \(Z\)
according to Appendix B.1 of Ref.~\cite{FBQC}. This implies that there are $2^6$ different possible resource states. We reproduce this behavior by sampling randomly the bias of every Shor code, with the constraint that two Shor codes that will be fused together must have the same bias. We build $n_{seed}$ noiseless fusion networks with different configurations in order to reflect the fusion network in general rather than a single bias configuration.

Every physical fusion is assigned an \(X\) or \(Z\) bias failure basis randomly, which will set the fusion error-model bias.
Detectors of the encoded fusion-network are set to ensure matchability of an agnostic error model, by ensuring no measurement outcome is involved in more than 2 detectors. Table~\ref{tab:detector-assignment} describes how the four measurements of each type, $m^{XX}_i$ and $m^{ZZ}_i$ ($i = 1, \dots, 4$) are assigned to the detectors. For $X$ ($Z$) measurements, there are two fusion network detectors $N_X^{\pm}$ ($N_Z^{\pm}$), the same as in the unencoded fusion network, and $1$ or $2$ detectors coming from the Shor-code, $(S_1,S_2)$ or $S_3$, depending on the Shor code bias.

\begin{table}[ht]
  \centering
  \small
    \centering
    \begin{tabular}{|c|c|c|}
      \hline
      Outcome & Z-type blocks& X-type blocks \\ \hline\hline
      $m^{XX}_1$ & $N_X^+$, $S_3$ & $N_X^+$, $S_1$ \\
      $m^{XX}_2$ & $N_X^-$, $S_3$ & $N_X^+$, $S_2$ \\
      $m^{XX}_3$ & $N_X^+$, $S_3$ & $N_X^-$, $S_1$ \\
      $m^{XX}_4$ & $N_X^-$, $S_3$ & $N_X^-$, $S_2$ \\
      $m^{ZZ}_1$ & $N_Z^+$, $S_1$ & $N_Z^+$, $S_3$ \\
      $m^{ZZ}_2$ & $N_Z^+$, $S_2$ & $N_Z^-$, $S_3$ \\
      $m^{ZZ}_3$ & $N_Z^-$, $S_1$ & $N_Z^+$, $S_3$ \\
      $m^{ZZ}_4$ & $N_Z^-$, $S_2$ & $N_Z^-$, $S_3$ \\\hline
    \end{tabular}

  \caption{Measurement outcome assignment to detectors.
  $N_X^\pm$ ($N_Z^\pm$): the network detectors containing the encoded $\overline{XX}$
  ($\overline{ZZ}$) outcome, on either side of the fusion. $S_1$, $S_2$, $S_3$: the Shor code
  detectors of the fusion.}
  \label{tab:detector-assignment}
\end{table}

\subsubsection{Sampling Noise and Randomness.} 
Randomness is sampled at two nested levels for the six ring, and three levels for the Shor-encoded six ring.
\begin{enumerate}
    \item For the Shor-encoded only, a random \emph{seed} fixes the bias of the Shor codes in the fusion network.
\item For a fixed Shor-bias layout, we sample a \emph{run}, in which all heralded events are drawn once and held fixed: erasures of the fusion outcomes, arising from the non-deterministic failure of the fusions. Erasures are inserted as errors on the circuit.
\item For a fixed run, many \emph{shots} are drawn, in which the circuit's detector error model is sampled. The detector error model accounts for the resource-state Pauli errors, the undetected fusion errors stemming from photon distinguishability, and the errors that have been added upon heralding. 
\end{enumerate}
This hierarchy is motivated by the fact that the \emph{Stim} circuits~\cite{gidney2021stim} requires the heralded erasure locations to be known before a noisy circuit can be sampled, and that constructing a new network per sample is very  costly in computation time. The number of seeds, run, and shots for the different simulations are given in Appendix~\ref{app:simulation}.

\subsubsection{Decoding and recording logical errors} For each shot, the outcomes are decoded with minimum-weight perfect matching using \emph{PyMatching} \cite{higgott2022pymatching}. 

Circuits' outcomes are sampled from the exact detector error models, but decoded with a modified one, in which correlated resource-state errors are split into smaller error components to avoid any non-graphlike errors, as minimum-weight perfect matching requires.
Because shots within a run share a heralded pattern, and runs within a seed share a Shor bias configuration, the shots are not independent. 

Each seed $s$ runs $H$ heralding runs of $n_{shot}$ shots each. The logical error rate of a seed $\epsilon_L^{(s)} $ is the fraction of failed shots  $E_s$ out of $N= n_{shot}H$ shots,  $\epsilon_L^{(s)} = E_s/N$, and the total logical error rate is
\begin{equation}
  \epsilon_L =  \frac{1}{n_{seed}}\sum_{s}{\epsilon_L^{(s)}}.
\end{equation}  
The uncertainty of $\epsilon_L$ is the standard error of the seed estimates,
\begin{equation}
  \sigma^2 = \frac{1}{n_{seed}(n_{seed}-1)} \sum_s \left(\epsilon_L^{(s)} - \epsilon_L\right)^2,\label{eq:sigma_6rs22}
\end{equation}

The unencoded network has a single layout; its uncertainty counts 
\begin{equation}\sigma^2 = \epsilon_L(1-\epsilon_L)/H. \label{eq:sigma_6r}\end{equation}

\subsection{Results}

\begin{figure*}[ht]
    \centering
    \includegraphics[width=0.95\linewidth]{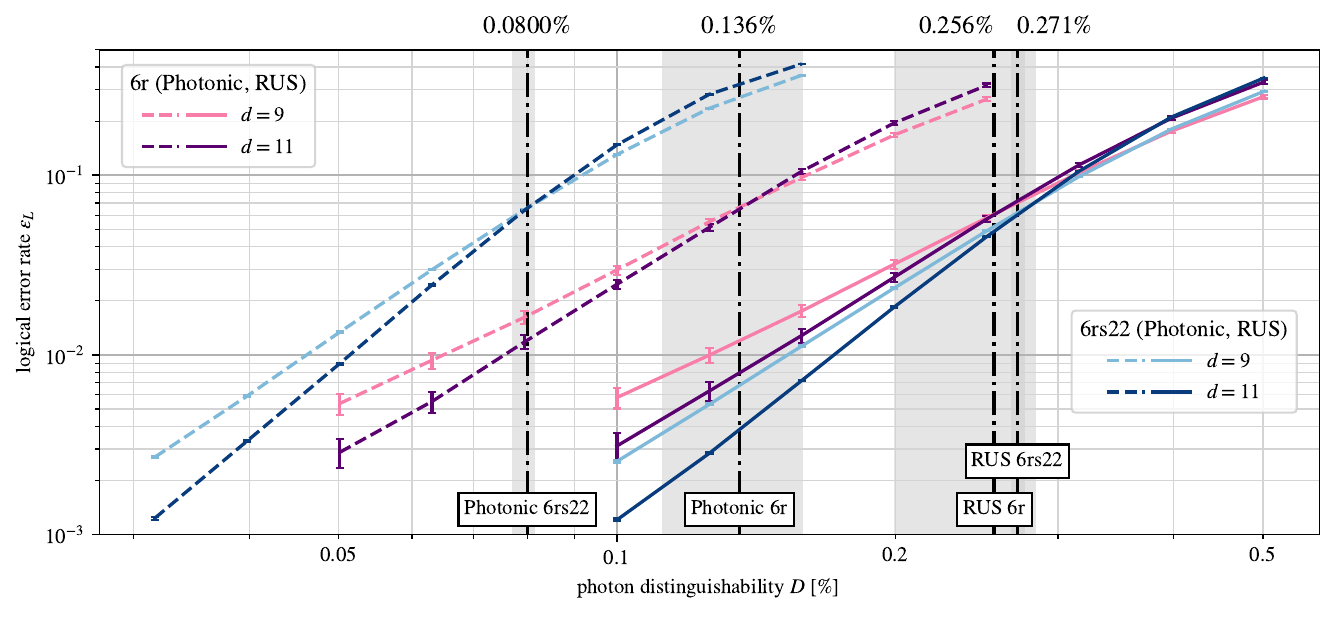}
    % \caption{\textcolor{red}{TEMPORARY.}}
\caption{
Logical error rate \(\epsilon_L\) as a function of photon distinguishability \(D\) for the unencoded and Shor$(2,2)$-encoded six-ring fusion networks, with all-photonic or emitter-assisted RUS resource preparation. Results are shown for code distances \(d=9\) and \(d=11\). Error bars indicate one standard deviation (1$\sigma$).
The simulations include distinguishability-induced noise during resource preparation and fusion measurements. 
Thresholds values are reported on the top of the plots and marked by the black dashed lines. $95\%$ confidence intervals of the thresholds are shaded in light gray. The derivation of the error bars and confidence intervals are explained in Appendix~\ref{app:simulation}.
}
\label{fig:distinguishability-thresholds}
    \label{fig:dist_thresholds}
\end{figure*}

Figure~\ref{fig:distinguishability-thresholds} shows the logical error rate $\epsilon_L$ as a function of photon distinguishability D for the unencoded and Shor$(2,2)$-encoded six-ring networks, using either all-photonic or emitter-assisted RUS resource preparation. For each configuration, we compare the code distances d=9 and d=11. 
We determine the threshold from the difference between the logical error rate at $d=9$ and $d=11$. At the simulated distinguishability, we find the two consecutive distinguishability points between which this difference changes sign, and interpolate it linearly in $\log D$ between them to find where it equals zero.
For all-photonic preparation, the estimated threshold decreases from $D^{\mathrm{th}}=0.136\%$ for the unencoded network to $D^{\mathrm{th}}=0.08\%$ for the encoded network. In contrast, the RUS construction gives comparable unencoded and encoded thresholds of \(0.256\%\) and \(0.271\%\), respectively.
Within the noise model considered here, the RUS construction therefore tolerates more distinguishability than the all-photonic construction for both networks. The effect of encoding depends on the preparation scheme and the associated fusion implementation: the unencoded network uses \(p_{\mathrm{fail}}=1/8\), whereas the encoded network uses \(p_{\mathrm{fail}}=1/4\).

\section{Discussion}
\label{sec:Discussion}

Our results show that the effect of local encoding on
distinguishability tolerance depends on how the resource states
are prepared. Two competing effects explain this. Encoding provides
redundancy that helps recover parity information erased by
photon loss or intrinsic fusion failure. In our configuration,
it also permits less strongly boosted physical fusions,
with an ideal failure probability of \(1/4\), compared with
\(1/8\) for the unencoded network. Preparing the larger encoded
resource state, however, requires additional operations that
can introduce distinguishability errors. The threshold
therefore reflects the combined effects of resource preparation,
encoding, and fusion implementation. For the all-photonic
construction studied here, the additional preparation noise
outweighs these benefits.

The structure of this noise is important. Heralded failures
identify missing parity outcomes, allowing the decoder to
exploit the remaining information. Partial photon
distinguishability can produce accepted detection patterns
with incorrect parity outcomes, additional heralded erasures
of one or both outcomes, and correlated errors during resource
preparation. These results complement the loss and fusion-failure analysis
of Ref.~\cite{FBQC}, showing that physical
implementations must also account for distinguishability-induced
noise during resource preparation and fusion.

The two preparation schemes produce different effective error
channels despite generating identical ideal resource states.
The all-photonic schemes combine noisy GHZ seeds through
successive fusion and post-selection stages, while the RUS
schemes establish connectivity through heralded entangling
operations between emitters. Our results demonstrate that
these preparation-dependent channels matter at the
fusion-network level. This complements recent work resolving
physical errors in photonic seed-state
generation~\cite{Bartolucci2026BeyondFidelity} and motivates
architecture benchmarks that propagate hardware-derived
noise through the actual preparation protocol.

In terms of hardware feasibility, heralded single photon sources have demonstrated indistinguishability between photons from different sources of up to 99.5\% \cite{psiquantum2025manufacturable} compared to the threshold of 99.92\% for the encoded all-photonic scheme, representing a 6-fold gap in error rate. Solid-state quantum dots have demonstrated up to 92\% indistinguishability between photons from separate GaAs quantum dots \cite{zhai2022quantum} compared to the threshold of 99.74\%, representing a 30-fold gap in error rate. However, recent measurements of low-charge noise InGaAs devices predict values up to 98\% can be achieved using high-Q micropillar cavities \cite{huet2606industry}, which if achieved would still imply an 8-fold gap in error. Furthermore, reaching threshold is only the minimum requirement for fault tolerance. Implementing utility-scale algorithms are expected to require logical error rates less than $10^{-12}$ and must tolerate multiple sources of error, which implies that individual error mechanisms must be significantly below their thresholds, typically an order of magnitude below. Such a requirement further increases the gap to almost two orders of magnitude relative to the state-of-the-art hardware. Hence, our study highlights a significant gap between hardware performance and fault tolerance requirements for practical fusion-based quantum computing.

Recent studies of fault-tolerant quantum-emitter-based architectures report distinguishability thresholds an order of magnitude higher than the thresholds obtained here, typically between 2\% and 4\%~\cite{SPOQC,dessertaine2024enhanced,chan2026practical,Chan2025}. Such thresholds are within reach of near-term improvements for quantum dot emitters. Using emitter spins to generate and process entanglement can reduce reliance on successive linear-optical interference and post-selection stages, offering a promising route to improved robustness against errors. These findings motivate exploiting emitter spins throughout the architecture, beyond their role as photon sources or resource state generators, and jointly optimizing the physical operations and error-correcting code. However, identifying the most advantageous implementation requires
assessing these gains together with photon loss tolerance, spin decoherence errors, temporal and spatial resource costs, and feasible hardware targets.

The present comparison remains specific to the constructions
studied and their modeling assumptions. We use first-order OBB noise channels with an effective Pauli description in the logical subspace, and we consider only the six-ring resource state with the Shor-$(2,2)$ encoding. Future work
could include leakage, use heralding information in post-selection
and decoding, and treat higher-order effects such as multiple
distinguishable photons and correlations across successive
interference stages. Distinguishability should also be combined
with photon loss, and other resource states, encodings, and
adaptive fusion strategies compared under consistent physical
assumptions. Finally, threshold estimates should be complemented
by logical error rates below threshold and resource
overhead~\cite{lobl2026subthreshold}.

\section{Conclusion and Outlook}
\label{sec:Conclusion}
We have studied how partial photon distinguishability affects
fusion-based fault-tolerant quantum computation when resource states are generated using realistic linear-optical preparation circuits. Most threshold estimations for fusion-based architectures assume that resource states are generated perfectly, with noise arising exclusively from subsequent fusion operations. Under this simplifying assumption, increasing the size of the resource states using logical encoding generally leads to improved error-correction performance. However, the generation protocols used to prepare these states are usually not fault-tolerant, meaning that errors can accumulate and propagate throughout the preparation process.

In the case of photon loss, these errors are heralded and can therefore be handled through post-selection. Although this reduces the resource generation rate, it does not fundamentally prevent the architecture from benefiting from the improved loss tolerance offered by larger encoded resource states. By contrast, unheralded errors, such as those arising from photon distinguishability, cannot be identified and discarded through post-selection. They therefore introduce intrinsic noise into the resource states, which can propagate through subsequent fusion operations and potentially offset the benefits of increasing the encoding size. We  have shown here that the error correction performances are actually reduced when using Shor(2,2) encoded resource states, contrary to what happens for loss correction performances.

Alternative resource-state preparation circuits offer promising opportunities to reduce preparation noise. Recent work develops a framework for resolving the sensitivity of seed-state generators to distinct physical error mechanisms, providing a basis for optimizing their noise performance~\cite{Bartolucci2026BeyondFidelity}. Another direction is the use of multi-way fusions during preparation. Bartolucci's analysis shows that these can reduce resource requirements and preparation-induced Pauli errors compared with constructions based exclusively on two-way fusions~\cite{Bartolucci2025Thesis}. For fusions between resource states in the computational network, however, that study finds evidence that two-way fusions remain preferable because of their lower erasure rate. Incorporating these alternative preparation circuits into our analysis would help determine how much the distinguishability thresholds can be improved and at what resource cost.

Adaptive fusion strategies and scalable post-selection provide promising directions for future work. Dynamic bias arrangement selects fusion failure bases according to previously observed erasures, suppressing the formation of large connected erasure clusters and improving photon-loss tolerance under the assumptions of Ref.~\cite{bombin2023increasing}. For our noise channels, changing the failure basis also interchanges the roles of $XX$ and $ZZ$ in the parity-error distribution. Assessing the benefit of this strategy therefore requires jointly accounting for erasures and distinguishability-induced errors. A complementary approach is Macromux, which uses hierarchical post-selection over constant-size spacetime regions to improve tolerance to both Pauli and erasure errors at additional resource cost~\cite{Macromux}. Evaluating these approaches with the noise models derived here would establish their impact on distinguishability thresholds and the associated resource overhead.

Our results establish resource-state preparation as an essential part of the fault-tolerance analysis. Comparisons of photonic architectures should account for how physical noise propagates through preparation circuits and affects the subsequent fusion outcomes and decoding. Extending this analysis to other resource states and encoding, and to both photon loss and distinguishability would provide a more complete assessment of architectural performance.

% add Paul Fundings
\section{ACKNOWLEDGMENTS}
The authors thank \textcolor{black}{Harshit Verma, Théo Dessertaine}  for fruitful discussions; \textcolor{black}{Boris Bourdoncle, Ellen Derbyshire} for valuable feed-back on the manuscript. This work has been co-funded by the Horizon CL4 program under the grant agreement n°101135288 for the EPIQUE project, and by the French National Quantum Strategy (France 2030) through the PROQCIMA and TUF-TOPIQC programs and the OECQ project. KH is supported by the CIFRE grant n°2025/0660. NS is supported by the National Research Foundation, Singapore, through the National Quantum Office, hosted in A*STAR, under its Center for Quantum Technologies Funding Initiative (S24Q2d0009). In addition, NS acknowledges support by the French National Quantum Strategy (France 2030) through the BACQ project as part of the MetriQs-France program (Grant ANR-22-QMET-0002).

\section{Author contributions}
K.H. developed the noise models, prepared the figures, and led the writing of the manuscript. G.G. proposed the project and, together with N.S., contributed to the simulations and manuscript writing. S.W. and P.H. supervised the work. All authors reviewed and edited the manuscript.

\appendix
\section{Boosted-fusion noise models}

In this appendix, we derive the distinguishability-induced error
channels and erasure probabilities for the two boosted Type-II
fusion schemes considered in the main text. We use the OBB model
with \(D=1-V_{\mathrm{HOM}}\). The calculations are conditioned on
no photon loss and assume ideal balanced beam splitters and
photon-number-resolving detectors that do not resolve the internal
modes.

\paragraph{Bell inputs and distinguishability model.}
In the four spatial modes of the two dual-rail input qubits,
the Bell states are
\begin{align}
\ket{\psi^\pm}
&=\frac{\ket{1001}\pm\ket{0110}}{\sqrt2},
&
\ket{\phi^\pm}
&=\frac{\ket{1010}\pm\ket{0101}}{\sqrt2}.
\label{eq:boosted-bell-states}
\end{align}

For a Bell input \(\beta\) and a total of \(N\) photons, including
the ancillary photons, Eq.~\eqref{eq:obb-input-state} gives the
equivalent input mixture
\begin{equation}
\rho_{\mathrm{in}}^\beta
=
\left(1-\frac{ND}{2}\right)\rho_0^\beta
+
\frac{D}{2}\sum_{j=1}^{N}\rho_j^\beta
+
\mathcal O(D^2).
\label{eq:boosted-input-noise}
\end{equation}
The state \(\rho_0^\beta\) contains the Bell input and ancillary
resources with all photons in internal mode \(0\).
In \(\rho_j^\beta\), photon \(j\) occupies its orthogonal internal
mode \(j\), while all other photons remain in mode \(0\).
Because the interferometer preserves the internal modes and the
detectors do not resolve them, we propagate each configuration
separately and sum the detection probabilities over the unresolved
internal modes.

\paragraph{Detection-outcome classification.}
We classify each detection record using the ideal interferometer's
Bell state discrimination rules. The same classification is used
for both boosting levels and distinguishes three outcome classes.

\textbf{Successful fusion.}
A pattern that uniquely identifies a Bell state in the ideal
measurement is accepted and reports both \(XX\) and \(ZZ\).
With partial distinguishability, the same pattern can occur for
a different Bell input. The decoder retains its ideal assignment,
so one or both reported parities may be incorrect. These errors
are unheralded.

\textbf{Partial erasure.}
The intrinsic failure patterns do not distinguish
\(\phi^+\) from \(\phi^-\), so they erase \(XX\) while retaining
\(ZZ\). These patterns occur even for indistinguishable photons.
In the first-order calculations below, the retained \(ZZ\)
outcome remains correct, although the probability of this
outcome class changes with \(D\).

\textbf{Full erasure.}
Partial distinguishability also produces patterns that have
zero probability for every Bell input in the ideal interferometer.
Some of these additional patterns can arise from all four Bell
states, preventing either parity from being assigned with certainty, so both outcomes are erased. These events occur even
when every photon is detected and therefore constitute
distinguishability-induced measurement erasures without photon loss.

\paragraph{Conditional error channel.}
For a Bell input \(\ket{\beta}\), we represent errors in the
reported Bell label by an effective Pauli operator
\(P\in\{I,X,Y,Z\}\) acting on one fusion input before an ideal
Bell measurement. We define
\begin{equation}
w_P^\beta=\Pr(\mathrm{s},P\mid\beta),
\qquad
p_{\mathrm{s}}^\beta
=\sum_{P\in\{I,X,Y,Z\}}w_P^\beta,
\label{eq:bell-success-probability}
\end{equation}
where \(\mathrm{s}\) denotes a successful fusion.
For equiprobable Bell inputs, we average these joint probabilities
over the four Bell states and normalize by the total success
probability to obtain the effective conditional Pauli channel.
The partial- and full-erasure probabilities are reported separately.

The following sections give these results for each boosting level,
quantifying how the reduction in ideal intrinsic failure competes
with additional distinguishability-induced errors and erasures.

\subsection{First-level Boosted-fusion noise model}
\label{app:boosted-fusion-noise}
The purpose of this section is to derive the effect of partial photon distinguishability on the boosted fusion scheme introduced in Ref.~\cite{3/4BSM}. 

\begin{figure}[H]
    \centering

    \resizebox{0.98\columnwidth}{!}{%
    \begin{tikzpicture}[
        x=0.52cm,
        y=0.55cm,
        font=\scriptsize,
        optical/.style={
            draw={rgb,255:red,145;green,12;blue,16},
            line width=1.05pt,
            line cap=round,
            line join=round
        },
        beamsplitter/.style={
            rectangle,
            draw=black!80,
            fill=gray!55,
            minimum width=5.5mm,
            minimum height=3.5mm,
            inner sep=0pt,
            line width=0.7pt,
            font=\tiny
        },
        detector/.style={
            rectangle,
            rounded corners=1.8pt,
            draw=purple!65!black,
            fill=purple!18,
            minimum width=6.5mm,
            minimum height=4.2mm,
            inner sep=0.5pt,
            line width=0.7pt,
            font=\tiny
        },
        input/.style={
            draw=none,
            anchor=east,
            inner sep=1pt
        }
    ]

    % Optical coupler
    \newcommand{\BS}[4]{%
        \pgfmathsetmacro{\ymiddle}{(#2+#3)/2}
        \draw[optical]
            ({#1-0.45},#2)
            -- (#1,\ymiddle)
            -- ({#1+0.45},#2);
        \draw[optical]
            ({#1-0.45},#3)
            -- (#1,\ymiddle)
            -- ({#1+0.45},#3);
        \node[beamsplitter] at (#1,\ymiddle) {\(#4\)};
    }

    % ========================================================
    % Inputs
    % ========================================================

    % Upper ancillary single photons
    \node[input] at (-0.15, 3.5) {\(\ket{1}_{a_1}\)};
    \node[input] at (-0.15, 2.5) {\(\ket{1}_{a_2}\)};

    % Fusion-input modes
    \node[input] at (-0.15, 1.5) {\(A\)};
    \node[input] at (-0.15, 0.5) {\(B\)};
    \node[input] at (-0.15,-0.5) {\(C\)};
    \node[input] at (-0.15,-1.5) {\(D\)};

    % Lower ancillary single photons
    \node[input] at (-0.15,-2.5) {\(\ket{1}_{a_3}\)};
    \node[input] at (-0.15,-3.5) {\(\ket{1}_{a_4}\)};

    % Brace grouping the four input modes
    \draw[
        decorate,
        decoration={brace,amplitude=4pt},
        gray!80!black,
        line width=0.7pt
    ]
    (-0.70,-1.75) -- (-0.70,1.75)
    node[
        midway,
        left=5pt,
        draw=none,
        align=right,
        text=black
    ] {
        fusion input\\[-1pt]
        \(\ket{\psi_1}\ket{\psi_2}\)
    };

    % ========================================================
    % First coupler layer
    % ========================================================

    % Ancilla preparation
    \draw[optical] (0, 3.5) -- (0.85, 3.5);
    \draw[optical] (0, 2.5) -- (0.85, 2.5);
    \BS{1.30}{3.5}{2.5}{R}

    \draw[optical] (0,-2.5) -- (0.85,-2.5);
    \draw[optical] (0,-3.5) -- (0.85,-3.5);
    \BS{1.30}{-2.5}{-3.5}{R}

    % Fusion-input couplers
    \draw[optical] (0,1.5) -- (3.55,1.5);
    \draw[optical] (0,0.5) -- (3.55,0.5);
    \BS{4.00}{1.5}{0.5}{R}

    \draw[optical] (0,-0.5) -- (3.55,-0.5);
    \draw[optical] (0,-1.5) -- (3.55,-1.5);
    \BS{4.00}{-0.5}{-1.5}{R}

    % ========================================================
    % Mode permutations before the final couplers
    % ========================================================

    % Upper outer rail
    \draw[optical] (1.75,3.5) -- (7.05,3.5);

    % Upper permutation: modes 1 and 2
    \draw[optical]
        (1.75,2.5) -- (5.80,2.5)
        -- (6.60,1.5) -- (7.05,1.5);

    \draw[optical]
        (4.45,1.5) -- (5.80,1.5)
        -- (6.60,2.5) -- (7.05,2.5);

    % Central permutation: modes 3 and 4
    \draw[optical]
        (4.45,0.5) -- (5.20,0.5)
        -- (6.00,-0.5) -- (7.05,-0.5);

    \draw[optical]
        (4.45,-0.5) -- (5.20,-0.5)
        -- (6.00,0.5) -- (7.05,0.5);

    % Lower permutation: modes 5 and 6
    \draw[optical]
        (4.45,-1.5) -- (5.80,-1.5)
        -- (6.60,-2.5) -- (7.05,-2.5);

    \draw[optical]
        (1.75,-2.5) -- (5.80,-2.5)
        -- (6.60,-1.5) -- (7.05,-1.5);

    % Lower outer rail
    \draw[optical] (1.75,-3.5) -- (7.05,-3.5);

    % ========================================================
    % Aligned final coupler layer
    % ========================================================

    \BS{7.50}{3.5}{2.5}{R}
    \BS{7.50}{1.5}{0.5}{R}
    \BS{7.50}{-0.5}{-1.5}{R}
    \BS{7.50}{-2.5}{-3.5}{R}

    % ========================================================
    % Aligned output permutations
    % ========================================================

    % Upper permutation
    \draw[optical]
        (7.95,2.5) -- (8.70,2.5)
        -- (9.50,1.5);

    \draw[optical]
        (7.95,1.5) -- (8.70,1.5)
        -- (9.50,2.5);

    % Lower permutation
    \draw[optical]
        (7.95,-1.5) -- (8.70,-1.5)
        -- (9.50,-2.5);

    \draw[optical]
        (7.95,-2.5) -- (8.70,-2.5)
        -- (9.50,-1.5);

    % ========================================================
    % Output rails
    % ========================================================

    \draw[optical] (7.95,3.5)  -- (11.50,3.5);
    \draw[optical] (9.50,2.5)  -- (11.50,2.5);
    \draw[optical] (9.50,1.5)  -- (11.50,1.5);
    \draw[optical] (7.95,0.5)  -- (11.50,0.5);

    \draw[optical] (7.95,-0.5) -- (11.50,-0.5);
    \draw[optical] (9.50,-1.5) -- (11.50,-1.5);
    \draw[optical] (9.50,-2.5) -- (11.50,-2.5);
    \draw[optical] (7.95,-3.5) -- (11.50,-3.5);

    % ========================================================
    % PNR detectors
    % ========================================================

    \foreach \y/\i in {
         3.5/0,
         2.5/1,
         1.5/2,
         0.5/3,
        -0.5/4,
        -1.5/5,
        -2.5/6,
        -3.5/7
    }{
        \node[detector,anchor=west] (D\i) at (11.80,\y)
            {\(D_{\i}\)};
        \draw[optical] (11.50,\y) -- (D\i.west);
    }

    % Labels
    \node[
        draw=none,
        text=gray!75!black,
        font=\tiny
    ] at (1.30,4.25) {ancilla preparation};

        draw=none,
        text=gray!75!black,
        font=\tiny
    ] at (12.15,4.25) {PNR detection};

    \end{tikzpicture}%
    }
\caption{Eight-mode interferometer implementing the boosted Type-II fusion scheme of Ref.~\cite{3/4BSM}. The balanced beam splitters are represented by the matrix \(R\) defined above. Modes \(A\)--\(D\) contain the two fusion inputs, while four ancillary single photons increase the ideal fusion success probability from \(1/2\) to \(3/4\). All output modes are measured using photon-number-resolving detectors.}
    \label{fig:boosted-typeII-interferometer}
\end{figure}
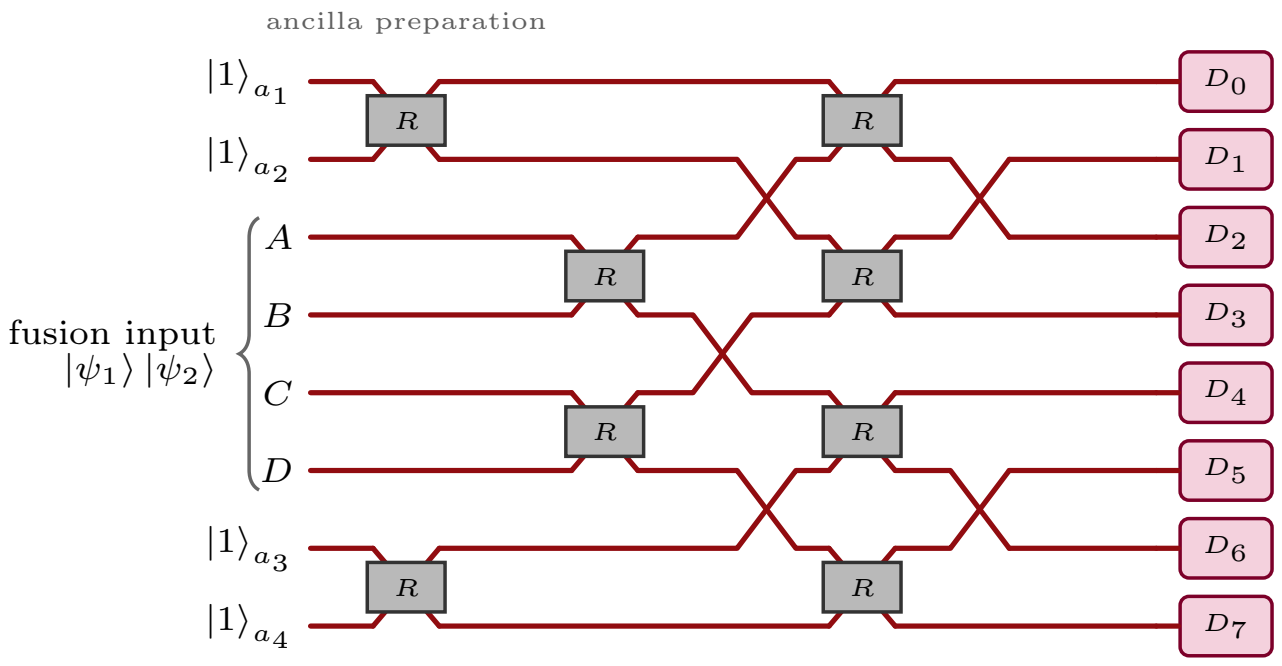

\paragraph{Optical interferometer}
The first-level scheme of Ref.~\cite{3/4BSM} uses eight spatial modes
and six photons. Modes \(2\)--\(5\) carry the two dual-rail input
qubits, and modes \(0,1,6,7\) contain four ancillary single photons.
Each balanced beam splitter marked \(R\) in
Fig.~\ref{fig:boosted-typeII-interferometer} implements
\begin{equation}
R=\frac1{\sqrt2}\begin{pmatrix}1&i\\i&1\end{pmatrix}.
\label{eq:balanced-bs-matrix}
\end{equation}
For indistinguishable ancillary photons, this transformation maps
\(\ket{11}\) to \(i\ket{\Upsilon_1}\), where
\begin{equation}
\ket{\Upsilon_1}=\frac{\ket{20}+\ket{02}}{\sqrt2}.
\label{eq:boosted-ancilla}
\end{equation}

The resulting two copies of \(\ket{\Upsilon_1}\) are used to assist the discrimination of the Bell states that remain ambiguous in the standard linear-optical Bell measurement.

The two ancillary pairs assist the discrimination of \(\phi^\pm\). Ideally, \(\psi^\pm\) are identified with probability one and \(\phi^\pm\) with probability \(1/2\), yielding the Bell-averaged success probability \(3/4\).

%------------------------------------------------------------
% Boosted Type-II fusion interferometer
%------------------------------------------------------------

\paragraph{Conclusive outcomes.}
Propagating the input configurations in Eq.~\eqref{eq:boosted-input-noise} through the interferometer gives the Bell-resolved  joint weights listed in Table~\ref{tab:bell-weights}. Each weight \(w_P^\beta\) is the probability of a conclusive outcome with effective Pauli error \(P\), conditioned on Bell input \(\beta\).

\begin{table}[ht]
\centering
\renewcommand{\arraystretch}{1.4}
\setlength{\tabcolsep}{6pt}
\begin{tabular}{c|cccc}
\hline
\(\beta\)
& \(w_I^\beta\)
& \(w_X^\beta\)
& \(w_Y^\beta\)
& \(w_Z^\beta\) \\
\hline
\(\phi^+\)
& \(\frac12-D\)
& \(\frac D2\)
& \(0\)
& \(\frac D4\) \\
\(\phi^-\)
& \(\frac12-D\)
& \(0\)
& \(\frac D2\)
& \(\frac D4\) \\
\(\psi^+\)
& \(1-\frac{3D}{2}\)
& \(\frac D4\)
& \(\frac D4\)
& \(\frac D2\) \\
\(\psi^-\)
& \(1-\frac{3D}{2}\)
& \(0\)
& \(0\)
& \(\frac D2\) \\
\hline
\end{tabular}
\caption{Bell-resolved joint probabilities of conclusive
outcomes and effective Pauli errors for first-level boosted
fusion.}
\label{tab:bell-weights}
\end{table}

\paragraph{Success and erasure probabilities.}
Summing the weights in each row and averaging over the four
equiprobable Bell inputs gives the total success probability.
Classifying the remaining detection records as partial or full
erasures gives
\begin{equation}
\begin{aligned}
p_{\mathrm s}^{(3/4)}
    &= \frac34-\frac D2+\mathcal O(D^2),\\
p_{\mathrm f}^{(3/4)}
    &= \frac14-\frac{3D}{16}+\mathcal O(D^2),\\
p_{\mathrm e}^{(3/4)}
    &= \frac{11D}{16}+\mathcal O(D^2).
\end{aligned}
\label{eq:boosted-success-failure-3-4}
\end{equation}
Here, \(p_{\mathrm f}\) includes all events where only \(XX\) is
erased, while \(p_{\mathrm e}\) describes erasure of both parities.

Averaging each column of Table~\ref{tab:bell-weights} over
the four Bell inputs and dividing by \(p_{\mathrm s}\)
gives the Pauli probabilities conditioned on success. We obtain
\begin{equation}
\begin{aligned}
\mathcal E_{\mathrm s}^{(3/4)}(\rho)
={}&(1-D)\rho
+\frac D4 X\rho X
+\frac D4 Y\rho Y\\
&+\frac D2 Z\rho Z
+\mathcal O(D^2).
\end{aligned}
\label{eq:app-boosted-3-4-channel}
\end{equation}
This error channel applies only to conclusive outcomes; the erasure probabilities are given separately in Eq.~\eqref{eq:boosted-success-failure-3-4}.

\subsection{Second-level boosted fusion noise model}
\label{app:boosted-fusion-noise-7-8}
This section extends the boosted-fusion calculation of Appendix~\ref{app:boosted-fusion-noise} from the first-level scheme, whose ideal success probability is 3/4, to the second construction of Ref.~\cite{3/4BSM}, whose ideal success probability is 7/8. 

The first-level scheme uses two copies of \(\ket{\Upsilon_1}\), each prepared from two single photons using a balanced beam splitter, as shown in Fig.~\ref{fig:boosted-typeII-interferometer}.
The second level adds two copies of the four-photon state
\begin{equation}
\ket{\Upsilon_2}
=\frac{\ket{2020}+\ket{0202}}{\sqrt2},
\end{equation}
one in each half of the interferometer. The optical interferometer follows the second-level construction shown in Fig.~2 of Appendix~D of Ref.~\cite{3/4BSM} and contains sixteen spatial modes and fourteen photons, including twelve ancillary photons.

We assume that each \(\ket{\Upsilon_2}\) resource is supplied in the intended coherent spatial superposition, with partial distinguishability as its only imperfection. Its preparation circuit and preparation success probability are not included in the model. The additional interference stage increases the ideal identification probability of each \(\phi^\pm\) input from \(1/2\) to \(3/4\), while \(\psi^\pm\) remain deterministically identified. For equiprobable Bell inputs, the ideal success probability is therefore \(7/8\).

\paragraph{Conclusive outcomes.}
We use Eq.~\eqref{eq:boosted-input-noise} with \(N=14\):
\begin{equation}
\rho_{\mathrm{in}}^\beta
=
(1-7D)\rho_0^\beta
+
\frac D2\sum_{j=1}^{14}\rho_j^\beta
+
\mathcal O(D^2).
\end{equation}
Propagating this state through the interferometer and applying the detection-outcome classification of Ref.~\cite{3/4BSM} gives the joint weights in Table~\ref{tab:bell-weights-7-8}.

\begin{table}[ht]
\centering

\renewcommand{\arraystretch}{1.4}
\setlength{\tabcolsep}{6pt}
\begin{tabular}{c|cccc}
\hline
\(\beta\)
& \(w_I^\beta\)
& \(w_X^\beta\)
& \(w_Y^\beta\)
& \(w_Z^\beta\) \\
\hline
\(\phi^+\)
& \(\frac34-\frac{45D}{16}\)
& \(\frac D2\)
& \(0\)
& \(\frac{15D}{16}\) \\
\(\phi^-\)
& \(\frac34-\frac{45D}{16}\)
& \(0\)
& \(\frac D2\)
& \(\frac{15D}{16}\) \\
\(\psi^+\)
& \(1-2D\)
& \(\frac D4\)
& \(\frac D4\)
& \(\frac D2\) \\
\(\psi^-\)
& \(1-2D\)
& \(0\)
& \(0\)
& \(\frac D2\) \\
\hline
\end{tabular}
\caption{Bell-resolved joint probabilities of conclusive outcomes and effective Pauli errors for second-level boosted fusion.}
\label{tab:bell-weights-7-8}
\end{table}

\paragraph{Success and erasure probabilities.}
Summing the weights in each row and averaging over the four equiprobable Bell inputs gives the total success probability. Classifying the remaining detection records as partial or full erasures gives
\begin{equation}
\begin{aligned}
p_{\mathrm s}^{(7/8)}
    &= \frac78-\frac{21D}{16}+\mathcal O(D^2),\\
p_{\mathrm f}^{(7/8)}
    &= \frac18+\frac{5D}{16}+\mathcal O(D^2),\\
p_{\mathrm e}^{(7/8)}
    &= D+\mathcal O(D^2).
\end{aligned}
\label{eq:boosted-success-failure-7-8}
\end{equation}

\paragraph{Conditional error channel.}
Averaging each column of Table~\ref{tab:bell-weights-7-8} over the four Bell inputs and dividing by \(p_{\mathrm s}\) gives the Pauli probabilities conditioned on success. Expanding to first order in \(D\), we obtain
\begin{equation}
\begin{aligned}
\mathcal E_{\mathrm s}^{(7/8)}(\rho)
={}&
\left(1-\frac{5D}{4}\right)\rho
+\frac{3D}{14}X\rho X\\
&+\frac{3D}{14}Y\rho Y
+\frac{23D}{28}Z\rho Z
+\mathcal O(D^2).
\end{aligned}
\label{eq:app-boosted-7-8-channel}
\end{equation}
This channel applies only to conclusive outcomes; partial and full erasures are accounted for separately in
Eq.~\eqref{eq:boosted-success-failure-7-8}.

\section{Distinguishability noise in GHZ-state preparation}
\label{app:ghz-noise}

We consider the three-photon and four-photon GHZ-state preparation
schemes based on primate states $\ket{\pi_n}$ introduced in
Ref.~\cite{Bartolucci2021Creation}. We first describe the
distinguishability model and conditioning assumptions common
to both constructions, then derive their effective logical
noise channels.

For \(n\) qubits, we define
\begin{equation}
\ket{\mathrm{GHZ}_n}
=
\frac{\ket{0}^{\otimes n}+\ket{1}^{\otimes n}}{\sqrt2},
\qquad
\rho_{\mathrm{GHZ}_n}
=
\ket{\mathrm{GHZ}_n}\bra{\mathrm{GHZ}_n}.
\end{equation}
Here, \(\ket{0}\) and \(\ket{1}\) denote dual-rail states.

We use the OBB model defined in the main text, with
\(D=1-V_{\mathrm{HOM}}\). For \(N\) input single photons,
the equivalent input mixture is
\begin{equation}
\rho_{\mathrm{in}}
=
\left(1-\frac{ND}{2}\right)\rho_0
+
\frac D2\sum_{j=1}^{N}\rho_j
+
\mathcal O(D^2).
\end{equation}
The state \(\rho_0\) contains all photons in internal mode
\(0\), while \(\rho_j\) contains photon \(j\) in its
orthogonal internal mode.

The interferometer preserves the internal modes. Its action
on the creation operators is therefore
\begin{equation}
\hat a_{r,\mu}^{\dagger}
\longrightarrow
\sum_m U_{mr}\hat a_{m,\mu}^{\dagger},
\end{equation}
where \(r,m\) label spatial modes and \(\mu\) labels the
internal mode. We propagate each input state separately and condition on the accepted detector patterns. The unresolved internal degrees of freedom are then traced out to obtain the state of the retained modes.

\paragraph{Leakage assumption.}
The accepted detection patterns contain exactly one photon in each detector pair. With distinguishability, however, such a record does not necessarily guarantee one photon in each retained dual-rail pair. The heralded output can therefore contain leakage outside the logical Hilbert space. In the noise model used here, we retain only the component with exactly one photon in each dual-rail pair and normalize this component. This is an optimistic assumption: some leakage events are not heralded, and excluding them would require additional information about the retained photons.
All results are expanded through first order in \(D\). The Pauli channels specify effective noise acting on the ideal GHZ resource state.

\subsection{Three-photon GHZ state}
\label{app:ghz3-noise}

The three-photon construction starts from six single photons and follows the sequence
\begin{equation}
\ket{1}^{\otimes6}
\longrightarrow
\ket{\pi_2}\otimes\ket{\pi_1}
\longrightarrow
\ket{\pi_3}
\longrightarrow
\ket{\mathrm{GHZ}_3}.
\end{equation}
The two-photon primate state is obtained by combining two
\(\ket{\pi_1}\) resources,
\(\ket{\pi_1}\otimes\ket{\pi_1}\longrightarrow\ket{\pi_2}\).

After conditioning on the accepted detector patterns and restricting to the logical subspace, the unnormalized
output is
\begin{equation}
\begin{aligned}
\widetilde{\rho}_{\mathrm{log}}^{(3)}
={}&
\frac{1-\frac94D}{32}\rho_{\mathrm{GHZ}_3}
+
\frac D{64}\sum_{i=1}^{3}
X_i\rho_{\mathrm{GHZ}_3}X_i\\
&+
\frac{3D}{256}
\left(
\ket{000}\bra{000}
+
\ket{111}\bra{111}
\right)
+
\mathcal O(D^2).
\end{aligned}
\label{eq:ghz3_obb_unnormalized}
\end{equation}
Using
\begin{equation}
\ket{000}\bra{000}+\ket{111}\bra{111}
=
\rho_{\mathrm{GHZ}_3}
+
Z_1\rho_{\mathrm{GHZ}_3}Z_1,
\end{equation}
and normalizing gives the effective logical channel
\begin{equation}
\begin{aligned}
\mathcal E^{(3)}(\rho)
={}&
\left(1-\frac{15D}{8}\right)\rho
+
\frac{3D}{8}Z_1\rho Z_1\\
&+
\frac D2\sum_{i=1}^{3}X_i\rho X_i.
\end{aligned}
\label{eq:ghz3_obb_channel}
\end{equation}

\subsection{Four-photon GHZ state}
\label{app:ghz4-noise}

The four-photon construction starts from eight single photons and follows the sequence
\begin{equation}
\ket{1}^{\otimes8}
\longrightarrow
\ket{\pi_2}\otimes\ket{\pi_2}
\longrightarrow
\ket{\pi_4}
\longrightarrow
\ket{\mathrm{GHZ}_4}.
\end{equation}

After conditioning on the accepted detector patterns and restricting to the logical subspace, the unnormalized
output is
\begin{equation}
\begin{aligned}
\widetilde{\rho}_{\mathrm{log}}^{(4)}
={}&
\frac{1-3D}{128}\rho_{\mathrm{GHZ}_4}
+
\frac D{512}\sum_{i=1}^{4}
X_i\rho_{\mathrm{GHZ}_4}X_i\\
&+
\frac D{512}
\Big(
4\ket{0000}\bra{0000}
+
\ket{0111}\bra{0111} +\ket{0100}\bra{0100}
\\
&+
\ket{0010}\bra{0010}
+
\ket{0001}\bra{0001}
\Big)
+
\mathcal O(D^2).
\end{aligned}
\label{eq:ghz4_obb_unnormalized}
\end{equation}
Unlike the three-photon result, this logical output is not a stochastic Pauli mixture acting on the ideal GHZ state.
To obtain a Pauli model, we apply the global bit-flip twirl
\begin{equation}
\mathcal T_X(\rho)
=
\frac12\left[
\rho+
X_1X_2X_3X_4\,\rho\,X_1X_2X_3X_4
\right].
\end{equation}

After twirling and normalization, the effective channel is
\begin{equation}
\begin{aligned}
\mathcal E^{(4)}(\rho)
={}&
\left(1-\frac{5D}{2}\right)\rho
+
\frac D2 Z_1\rho Z_1\\
&+
\frac{3D}{8}\sum_{i=1}^{4}X_i\rho X_i\\
&+
\frac D8\sum_{i=1}^{4}
Z_1X_i\,\rho\,X_iZ_1
+
\mathcal O(D^2).
\end{aligned}
\label{eq:ghz4_obb_channel}
\end{equation}

\section{Simulation settings and plots}
\label{app:simulation}
\paragraph{Simulation settings.} The simulation settings that lead to the threshold plot in Fig.~\ref{fig:dist_thresholds} are shared in Table~\ref{tab:sim-params}. 
We simulate both fusion networks, the unencoded and the Shor(2,2) encoded six-ring, each with resource states prepared by the all-photonic and the RUS schemes, which give four constructions. Their simulation parameters are listed in Table~\ref{tab:sim-params}.

\begin{table*}[t]
  \centering
  \setlength{\tabcolsep}{4pt}
  \begin{tabular}{|l|c|c|}\hline
    Parameter & Unencoded 6-ring & Shor-(2,2) encoded 6-ring \\ \hline \hline
    Fusion gate                       & boosted Ewert--van Loock & Ewert--van Loock \\
                                      & $p_\text{fail} = 1/8$, 14 photons & $p_\text{fail} = 1/4$, 6 photons \\
    Code distances                    & 9, 11 & 9, 11 \\
    Lattice seeds $n_{\text{seed}}$   & 1 (single layout) & 160 \\
    Heralding runs per task, $H$      & $10\,000$ & 128 \\
    Shots per heralding run, $n_{\text{shot}}$ & 128 & 128 \\
    Heralding runs per point (at most) $n_{\text{seed}}H$ & $10\,000$ & $20\,480$ \\
    Shots per point (at most) $n_{\text{shot}}n_{\text{seed}}H$ & $1.28 \times 10^{6}$ & $2.62 \times 10^{6}$ \\
    Error cap per task                & $1\,200\,n_{shot}$ & $10\,n_{shot}$ \\\hline
  \end{tabular}
  \caption{Simulation parameters. Each network uses the least boosted fusion with which it has a threshold according to Figure 10 of Ref.~\cite{FBQC}. The unencoded network has a single layout, so its statistics come from the heralding runs only; for the encoded network, the spread between layouts dominates, so its statistics come from the lattice seeds.}
  \label{tab:sim-params}
\end{table*}

\paragraph{Error bars.} Error bars are based on the standard deviation, see Eq.\eqref{eq:sigma_6rs22} and Eq.\eqref{eq:sigma_6r}. On the plot in Fig.~\ref{fig:dist_thresholds}, the higher bar is $\epsilon_L+\sigma$ and the lower bar is $\min{(\epsilon_L-\sigma,0.9\:\epsilon_L )}$.
\paragraph{Threshold confidence intervals.}
The $95\%$ confidence intervals of the six-ring networks are derived by drawing, 2000 times, every point of the $d = 9$ and $11$ curves independently from the normal distribution $\mathcal{N}(\epsilon_L, \sigma)$, with $\sigma$ given by Eq.\eqref{eq:sigma_6r}, and by taking the $2.5^{th}$ and $97.5^{th}$ percentiles of the 2000 resulting crossings.

For encoded six-ring, the same lattices are simulated at every distinguishability, so the points of a curve are correlated, and the resampling is done over lattices instead of points: for each distance, $n_{seed}$ seeds are drawn at random with replacement from the $n_{seed}$ simulated ones, so that some seeds appear several times and others not at all, and the logical error rate of every point is recomputed from this same set of seeds. The confidence interval is again given by the $2.5^{th}$ and $97.5^{th}$ percentiles of the 2000 resulting crossings.

\FloatBarrier 
\bibliographystyle{apsrev4-2}
\bibliography{biblio}

\end{document}